\documentclass[notitlepage,aps,pra,floatfix,superscriptaddress,showpacs,nofootinbib,longbibliography]{revtex4-1}
\usepackage{amsmath, amsthm, amssymb,amsfonts,mathbbol,amstext}
\usepackage{graphicx}
\usepackage{dcolumn}
\usepackage{bm}
\usepackage{bbm}
\usepackage{hyperref}
\usepackage{mathtools}
\usepackage{color}
\usepackage{multirow}
\usepackage{diagbox}
\usepackage{float}
\usepackage{xcolor}
\usepackage{listings}

\usepackage{changes}

\def\1{\mathbf{1}}
\def\0{\mathbf{0}}

\newcommand*{\coloneqq}{\mathrel{\vcenter{\baselineskip0.5ex \lineskiplimit0pt \hbox{\scriptsize.}\hbox{\scriptsize.}}} =}

\newcommand{\ket}[1]{| #1 \rangle}
\newcommand{\bra}[1]{\langle #1 |}

\newcommand{\processnext}[1]{%
	\ifx\listfinish#1\empty\else\listact{#1}\expandafter\processnext\fi}

\newcommand{\eqnref}[1]{(\ref{#1})}

\newcommand{\ignore}[1]{}
\newcommand{\nobibentry}[1]{{\let\nocite\ignore\bibentry{#1}}}

\newcommand{\bibfnamefont}[1]{#1}
\newcommand{\bibnamefont}[1]{#1}
\newcommand{\average}[1]{\left<#1\right>}
\let\oldsqrt\sqrt
\def\sqrt{\mathpalette\DHLhksqrt}
\def\DHLhksqrt#1#2{\setbox0=\hbox{$#1\oldsqrt{#2\,}$}\dimen0=\ht0
	\advance\dimen0-0.2\ht0
	\setbox2=\hbox{\vrule height\ht0 depth -\dimen0}%
	{\box0\lower0.4pt\box2}}

\begin{document}
	\title{Operational significance of nonclassicality in nonequilibrium Gaussian quantum thermometry	}
	\author{Safoura  Mirkhalaf}
	\email{s.mirkhalaf@cent.uw.edu.pl; SSM and MM contributed equally.}
	\address{Department of Physics, University of Tehran, P.O. Box 14395-547, Tehran, Iran}
	\address{School of Nano Science, Institute for Research in Fundamental Sciences (IPM), P.O. Box 19395-5531, Tehran, Iran}
	\address{Centre for Quantum Optical Technologies, Centre of New Technologies, University of Warsaw, Banacha 2c, 02-097 Warszawa, Poland}
	\author{Mohammad Mehboudi}
	\email{mohammad.mehboudi@tuwien.ac.at}
        \address{Technische Universität Wien, 1020 Vienna, Austria}
        \address{D\'epartement de Physique Appliqu\'ee, Universit\'e de Gen\`eve, 1211 Gen\`eve, Switzerland}
        \author{Zohre Nafari Qaleh}
        \address{School of Physics, Institute for Research in Fundamental Sciences (IPM), P.O. Box 19395-5531, Tehran, Iran}
	\author{Saleh Rahimi-Keshari}
	\email{keshari@ipm.ir}
	\address{Department of Physics, University of Tehran, P.O. Box 14395-547, Tehran, Iran}
	\address{School of Physics, Institute for Research in Fundamental Sciences (IPM), P.O. Box 19395-5531, Tehran, Iran}
	\begin{abstract}
  We provide new operational significance of nonclassicality in nonequilibrium temperature estimation of bosonic baths with Gaussian probe states and Gaussian dynamics. We find a bound on the thermometry performance using classical probe states. Then we show that by using nonclassical probe states, single-mode and two-mode squeezed vacuum states, one can profoundly improve the classical limit. Interestingly, we observe that this improvement can also be achieved by using Gaussian measurements. Hence, we propose a fully Gaussian protocol for enhanced thermometry, which can simply be realized and used in quantum optics platforms. 
	\end{abstract}
	\maketitle
	\section{Introduction}
	Temperature estimation is essential in characterizing and engineering quantum systems for any technological applications. Given finite experimental resources, quantum thermometry designs \textit{probes} with maximum information gain~\cite{Mehboudi_2019,de2018quantum}.  Along with the number of probes, \textit{time} is an important resource. Without any time restriction, it is often best to let a temperature probe thermalise with the system before reading it out. In this procedure, known as equilibrium thermometry, the initial state of the probe does not play a role and therefore the potential of quantum resources in the state preparation is not used. Nonetheless, the impact of initial probe states in a nonequilibrium (dynamical) scenario can be important and has been the subject of previous studies~\cite{PhysRevA.84.032105,PhysRevA.86.012125,PhysRevLett.114.220405,PhysRevA.91.012331,razavian2019quantum,PhysRevA.96.012316,PhysRevA.105.L030201,PhysRevA.98.050101,PhysRevA.92.052112,PhysRevLett.125.080402,PhysRevLett.123.180602,PhysRevA.102.042417,PhysRevResearch.2.033498,Henao2021catalytic,PRXQuantum.2.020322,Pavel_nonequilibrium}
	
	Continuous-variable systems in Gaussian states are the relevant model in many physical platforms for thermometry such as quantum optics, quantum gases, Josephson junctions, and mechanical resonators \cite{weedbrook2012gaussian}. In this context, a framework for using Gaussian measurements was recently proposed with some preliminary results at thermal equilibrium ~\cite{cenni2021thermometry}. On the other hand, in a recent paper~\cite{Pavel_nonequilibrium}, the limits of nonequilibrium thermometry in the Markovian environments were established. The results of~\cite{Pavel_nonequilibrium} are based on bounds in terms of operator norm that are only applicable to probe systems with a finite-dimensional Hilbert space. 
 Therefore, of particular interest is to understand the limits and advantages of 
 continuous-variable probe systems with infinite-dimensional Hilbert spaces---that are readily available in the laboratory 
  in Gaussian states---in nonequilibrium thermometry.
	
	In this paper, we propose a fully Gaussian setup with simple physical realizations that exploits the power of nonclassicality for nonequilibrium thermometry. Specifically, in our setup, the interaction between a continuous-variable probe system and a thermal bath is described by a class of Gaussian dynamics that can be described as a Brownian motion~
 \cite{PhysRevResearch.2.033498,PhysRevLett.114.220405,schaller2014open}. We first formalize the limit of classical probes for thermometry and show that this limit can be profoundly improved with the use of nonclassical Gaussian states and measurements. Our formalism provides a new insight into the operational significance of nonclassical states in nonequilibrium thermometry, and can readily be employed e.g., in quantum optical platforms.
	
    The structure of the paper is as follows. In Section~\ref{sec:setup} we 
    formalise the problem, the setup, and the figure of merit. In Section~\ref{sec:Gaussian_formalism}, we review the Gaussian formalism, including the Gaussian dynamics. 
    We then proceed with our main results in Section~\ref{sec:main_results}. Finally, in Section~\ref{sec:Discussion}, we close the paper with some remarks. Derivations and details of some results, as well as more simulations, are presented in the appendices in order to keep the paper's main part coherent.
	
	\section{The setup}\label{sec:setup}
\begin{figure}
		\centering
		\includegraphics[width=.5\columnwidth]{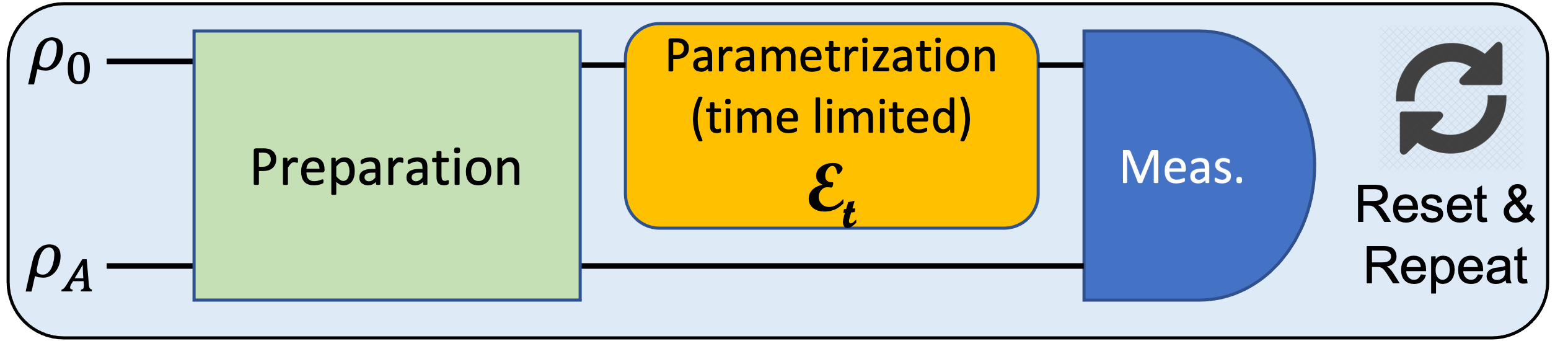}
		\caption{Our setting for estimating the temperature of a system: the probe is prepared in a state, which may or may not be correlated to auxiliary units, and then interacts with the sample for some time $t$. A measurement on the probe (and possibly the auxiliary units) will reveal some information about the temperature. One then resets the probe's state and repeats this process $M=\tau/t$ times, aiming to maximize the acquired information.}
		\label{fig:set_up}
	\end{figure}

 As shown in Fig.~\ref{fig:set_up}, the probe mode, which can be in an entangled state with an auxiliary mode, interacts with a bath at temperature $T$. After some interaction time $t$, the probe system is measured to infer the temperature of the bath. We assume that the total time of running the experiment is fixed to $\tau$. During this time, one can reset and repeat the whole process $M=\tau/t$ times. In what follows we assume that the time required for preparation and measurement protocols is not relevant compared to the time required for the parametrization (the probe's dynamics). Given that the density matrix of the system right after parametrization reads ${\cal E}_{t}(\rho)$, one performs a POVM measurement $\Pi$ with elements $\{\Pi_k\}$. After $M$ rounds, we can collect the outcomes to the dataset ${\bf x}=\{x_1,\dots,x_M\}$. An estimator function ${\tilde T}({\bf x})$ maps the dataset into an estimate of the true parameter. From the Cram\'er-Rao bound we know that the mean square error of any unbiased estimator is bounded from below \cite{cramer1999mathematical,newey1994large}
	\begin{align}\label{eq:CRB}
		{\rm MSE}({\tilde T};\rho;\Pi;t)\!\coloneqq\! \average{\!\big(\!{\tilde T}({\bf x}) - T_0\big)^{\!2}}_{\!\bf x}\!\!\!\geqslant\!\frac{t}{\tau {\cal F}^{\rm C}(\rho;\Pi;t)},
	\end{align}
	where the bound is saturable by choosing the maximum likelihood estimator. Here, $T_0$ is the mean value of the estimator over all  measurement outcomes, say $T_0=\langle\tilde{T}(x)\rangle_x$; ${\cal F}^{\rm C}(\rho;\Pi;t)$ is the classical Fisher information (CFI) associated with the measurement $\{\Pi\}$ and the initial state $\rho$. The CFI is upper bounded by the quantum Fisher information (QFI) defined as ${\cal F}^{\rm Q}(\rho;t)\coloneqq \max_{\Pi}{\cal F}^{\rm C}(\rho;\Pi;t)$~\cite{BRAUNSTEIN1996135,PhysRevLett.72.3439,paris2009quantum,giovannetti2011advances,hayashi2006quantum}. Motivated by \eqref{eq:CRB} and following~\cite{PhysRevLett.109.233601,PhysRevLett.114.220405,Pavel_nonequilibrium,PhysRevLett.110.050403} we set the rate of CFI, ${\tilde {\cal F}^{\rm C}}(\rho;\Pi;t)\coloneqq {\cal F}^{\rm C}(\rho;\Pi;t)/t$, and the rate of QFI, $\tilde{\cal F}^{\rm Q}(\rho;t)\coloneqq {\cal F}^{\rm Q}(\rho;t)/t$, as our figures of merit. There are three main variables to optimize over, the initial state $\rho$, the interrogation time $t$, and the measurement $\Pi$. 
	
	Our main aims are two folded (I) to understand the impact of Gaussian, nonclassical probe states and (II) to study the performance of Gaussian measurements. These are motivated by experimental feasibility; in quantum optics, preparation of Gaussian states is always possible via single-mode displacement and squeezing operations together with linear-optical networks, and Gaussian measurements are readily available via homodyne detection and a linear-optical network as well. We show that nonclassicality in Gaussian states, namely single-mode squeezing and two-mode squeezing can significantly improve the classical bound on thermometry precision. Moreover, Gaussian measurements often perform as well as the optimal measurement, especially in short time scales.
	
	\section{Gaussian formalism}\label{sec:Gaussian_formalism}
		We assume that probes are initially prepared in Gaussian states and their marginal states remain Gaussian through the interaction. Let us denote the vector of canonical operators by $R = (x_1,p_1,\dots, x_m,p_m)^T$, which also defines the matrix elements of the symplectic form $[R_j,R_k] \eqqcolon i\hbar \Omega_{jk} $. A Gaussian state is fully described by the vector of the mean values $d \coloneqq {\rm Tr}[\rho R]$ and the covariance matrix ($\hbar=1$)
	\begin{align}
		\sigma \coloneqq {\rm Tr}\!\left[\rho R\circ R^T\right] - 2d d^T,
	\end{align}
	where we use the notation $R\circ R^T \equiv RR^T + (RR^T)^T$ with $RR^T$ being a matrix of operators that is not symmetric.
	
	In the most general case, a Lindbladian Gaussian dynamics ${\cal E}_{t}$ evolves any probe state as \cite{demoen1977completely,Lindblad_2000,doi:10.1142/p489,PhysRevA.63.032312}
	\begin{gather}
		d_t = X_t d,\hspace{1cm}
		\sigma_t = X_t \sigma X^T_t + Y_t.
	\end{gather}
	Thus, the double $(X_t \in {\mathbb R}^{2m}\times{\mathbb R}^{2m} ,Y_t \in {\mathbb R}^{2m}\times{\mathbb R}^{2m})$ fully characterise the dynamics. They satisfy $Y_t^T=Y_t$ and $\Omega + Y + X_t \Omega X_t^T \geqslant 0$ to guarantee complete positivity of the dynamics. 
 
 For the standard Brownian motion the temperature information enters the dynamics only through the noise matrix $Y_t$ while the drift matrix $X_t$ is independent of temperature. This is the case in some situations such as the damped harmonic oscillator or loss of an optical cavity and has been previously used to study various problems in quantum thermometry~\cite{PhysRevResearch.2.033498,PhysRevLett.114.220405,schaller2014open}. In Appendix~\ref{app:Brownain}, we 
 provide a microscopic derivation of this master equation where we prove the term $X_t$ is indeed temperature independent. 
    More specifically, such Gaussian dynamical evolution can be described by $X_t = \exp(-\gamma t/2)O_t$, and $Y_t = (1-\exp(-\gamma t))\sigma_T$, where $O_t$ is an orthogonal matrix and $\sigma_T = \coth(\omega/2T)I_2$ is the covariance matrix of a bosonic mode with frequency $\omega$ at thermal equilibrium with temperature $T$. Here, $I_2$ is the $2\times2$ identity matrix. The parameter $\gamma=J(\omega)$ is related to the spectral density of the environment $J(\omega)$ and generally depends on the system's bare frequency $\omega$; however, it is independent of the environment temperature. Also, in our model, the orthogonal transformation $O_t$ corresponds to the coherent dynamics of the probe system that is independent of temperature. Hence, we can always work in the interaction picture with $O_t=I_2$ and incorporate the phase rotation in time into the measurement. This also implies that, without loss of generality, we can assume that the initial covariance matrix is diagonal, as  
	${\cal E}_{t}(U(\theta)\rho U^\dagger(\theta))=U(\theta){\cal E}_{t}(\rho)U^\dagger(\theta)$ 
	and the phase rotation $U(\theta)$ can further be absorbed into the measurement. In this case, the covariance matrix $\sigma_t$ remains diagonal and is given by
	\begin{equation}\label{eq:CM-time}
		\sigma_t=e^{-\gamma t} \sigma+ \big(1-e^{-\gamma t}\big)\nu I_2
	\end{equation}
	where $\nu=\coth(\omega/2T)$.
	
\begin{figure}
		\centering
		\includegraphics[width=.4\linewidth]{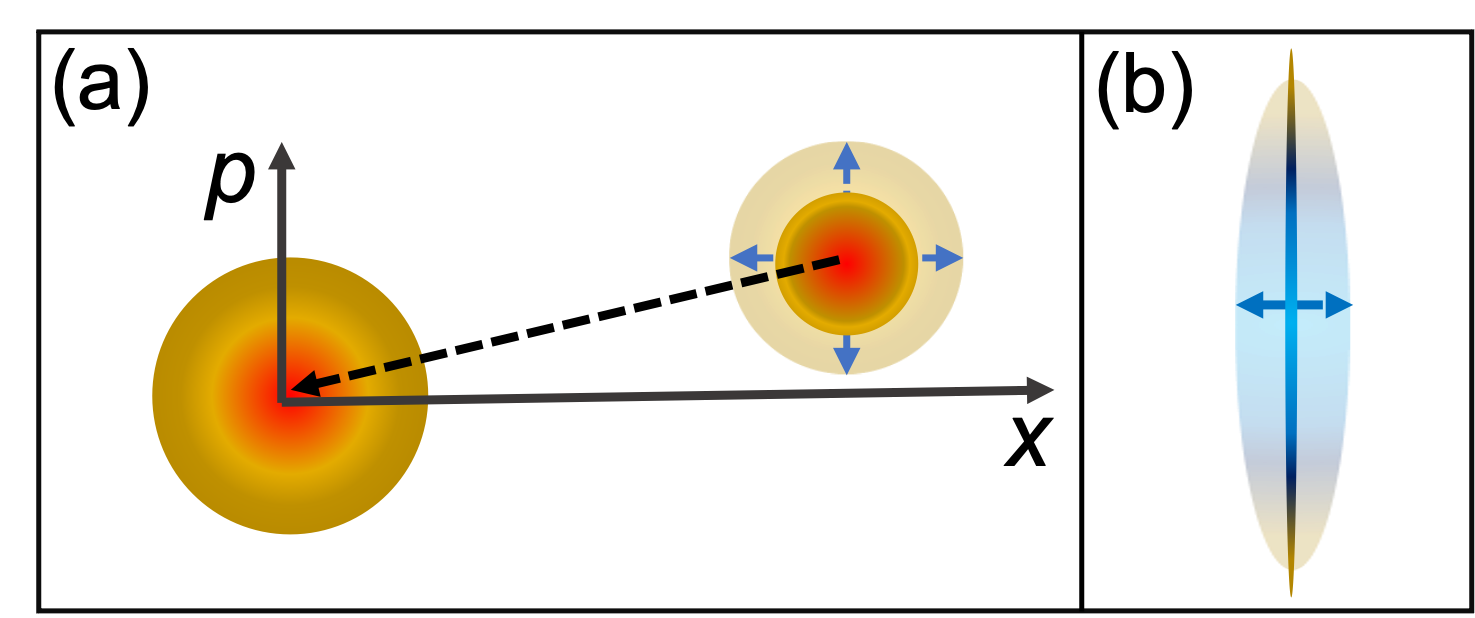}
		\caption{(a) Schematic representation of the probe's evolution in phase space. In our model, the quadrature mean values tend to zero (dashed black arrow) in a temperature-independent manner. The temperature rather determines how much noise is being added to the state (blue arrows). (b) If the probe is in a highly squeezed-vacuum state, the squeezed quadrature (narrow dark blue region) will be highly sensitive to temperature, while the temperature information imprinted in the conjugate quadrature is almost negligible compared with the initial large uncertainty. Therefore, homodyne measurement on the squeezed quadrature---which is a Gaussian measurement---is \textit{the optimal} measurement.}
		\label{fig:Phase_Space}
	\end{figure}

	Having the Gaussian state of the probe at the interrogation time $t$, one can calculate the QFI~\cite{monras2013phase,PhysRevA.88.040102,PhysRevA.89.032128,PhysRevA.98.012114} 
	\begin{align}
		\begin{split}\label{eq:QFI-Gaussian}
			{\cal F}^{\rm Q}(\sigma,d;t) & \!= 2\partial_T d_t^T \sigma_t^{-1}\partial_Td_t
        + \frac{1}{2} \bra{\partial_T\sigma_t}(\sigma_t\otimes\sigma_t - \Omega\otimes\Omega)^{-1}\ket{\partial_T\sigma_t},
		\end{split}
	\end{align}
	where $\partial_T$ denotes the partial derivative with respect to temperature, and we used the vectorization notation $\ket{\partial_T\sigma_t} \coloneqq ([\partial_T\sigma_t]_{11},\dots, [\partial_T\sigma_t]_{1m},\dots, [\partial_T\sigma_t]_{mm})^T$. Furthermore, if we perform a Gaussian measurement described by the covariance matrix $\sigma^M$, then the corresponding CFI reads~\cite{cenni2021thermometry,monras2013phase,malago2015information}
	\begin{align}\label{eq:CFI-Gaussian}
		{\cal F}^{\rm C}(\sigma,d;\sigma^M;t)=2\partial_T d_t^T (\sigma_t+\sigma^M)^{-1}\partial_Td_t
  +\frac{1}{2}{\rm Tr}\! \left[\left(({\sigma}_t + {\sigma}^{M})^{-1} \partial_{T}{\sigma}_t\right)^2 \right].
	\end{align}
However, as $X_t $ is temperature independent, no information about the temperature is imprinted on $d_t$; see Fig.~\ref{fig:Phase_Space}. Therefore, the first term in the above expressions for QFI and CFI vanishes.

	\section{Main results}\label{sec:main_results}
	
 Our main results are as follows. (i) The precision of using classical probe states is upper bounded by that of using the vacuum probe state.
	(ii) This bound can be overcome by using nonclassical states, namely single and two mode-squeezed states. In particular, we prove that single-mode squeezed states perform significantly better than the vacuum state, which previously was thought to be the best reparation~\cite{PhysRevLett.114.220405}. Moreover, entanglement between the probe and an auxiliary mode (two-mode squeezed states) further improves the thermometry precision. (iii) We prove the (extent of) optimality of Gaussian measurements, specifically homodyne detection for non-equilibrium thermometry. 
 \subsection{Result (i): 
	An upper bound for classical states}
	To begin with, we characterize the performance of classical states. A quantum state is classical if and only if its Glauber-Sudarshan $P$-function is a probability density distribution, and the state can be written as a mixture of coherent states, $\rho^{\rm cl}\coloneqq \int d^{2}\alpha P(\alpha)\ket{\alpha}\bra{\alpha}$~\cite{TituGlau1965,Mandel1986}. As discussed, in our model, no information about the temperature is imprinted on the first-order moments of the probe state. Therefore, the estimation precision of using all coherent probe states (including the vacuum probe state), which have the same covariance matrix $\sigma_{1}=I_2$, are the same. Also, as the Fisher information is convex---i.e., statistical mixture reduces the Fisher information---we conclude that the performance of all classical probe states is upper bounded by that of the vacuum probe state, that is,
	\begin{align} \label{eq:class-bound}
		{\cal F}^{\rm Q}(\rho^{\rm cl};t) \leqslant {\cal F}^{\rm Q}(\ket{\alpha}\bra{\alpha};t)={\cal F}^{\rm Q}(\ket{0}\bra{0};t), \hspace{.2cm} \forall ~t.
	\end{align}
	Note that this result holds for nonGaussian classical states as well.
	
	
\subsection{Result (ii): Beating the classical bound 
with nonclassical states}
\begin{figure}
		\centering
		\includegraphics[width=.3\columnwidth]{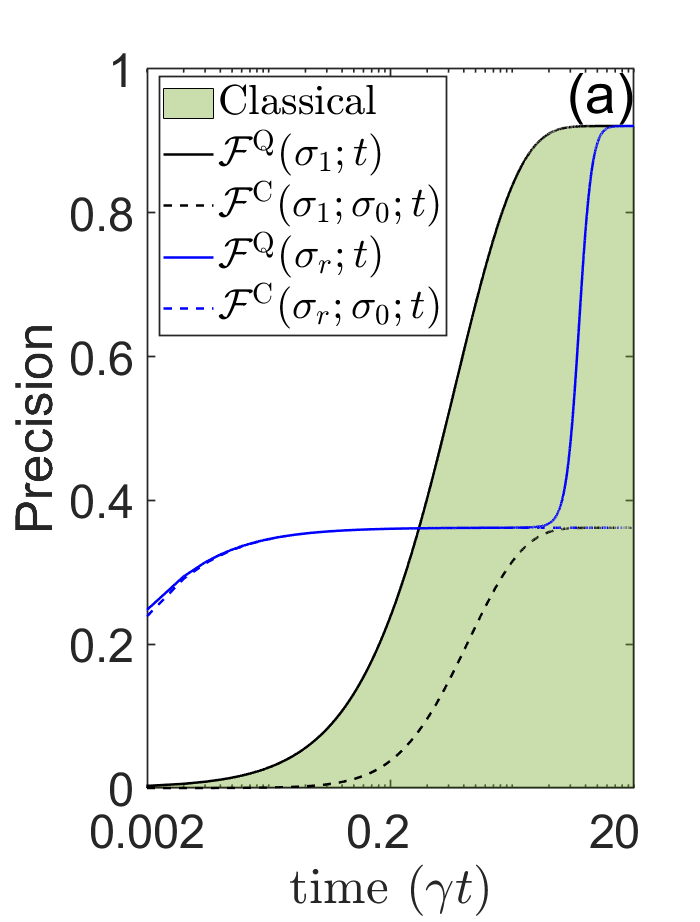}
		\includegraphics[width=.3\columnwidth]{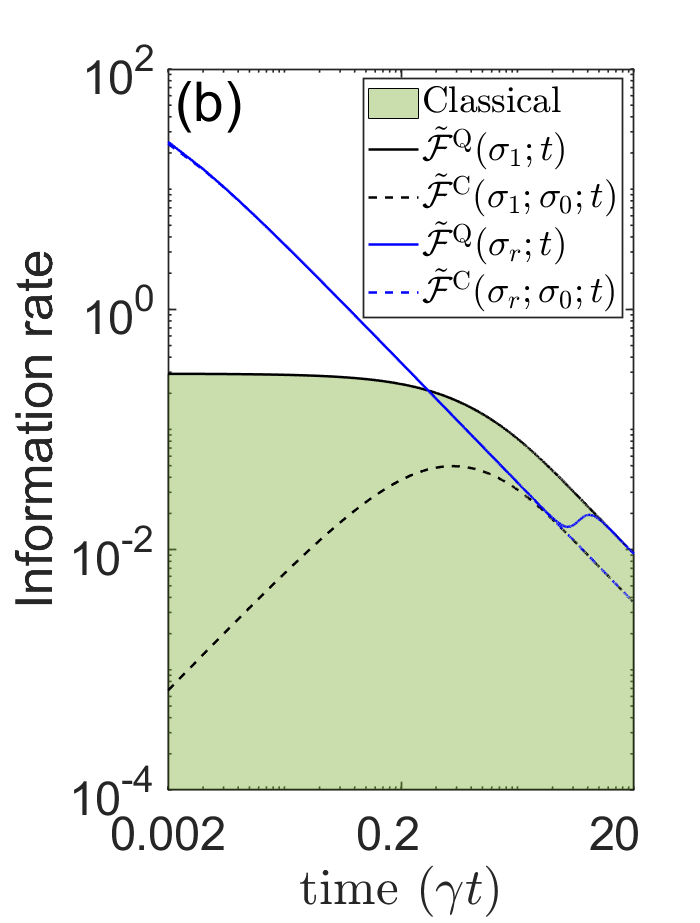}
		\caption{(a) The quantum Fisher information and (b) its rate are depicted in solid. The probe is initially prepared in the  vacuum or the squeezed state. The (rate of the) classical Fisher information of optimal Gaussian measurements are also plotted in dashed. All classical probe states	have a precision (or rate) that lies within the shaded green area, whose border is determined by the vacuum state. One can see that squeezed states can overcome this classical bound at short times. Interestingly, this improvement can still be attained by using Gaussian measurements, which are readily available in the laboratory. The optimal measurement should be performed as quickly as possible. As for the optimal Gaussian measurement, however, one may wait until the rate reaches its maximum, before measuring the system---better seen from Fig.~\ref{fig:QFI_rates_two_modes} that depicts short-time behavior.
			Here we set $T=1$, $\omega=1$, $\gamma =0.2$, $r =10^{-3}$.}
		\label{fig:QFI_rates_single}
	\end{figure}

As the first-order moments and a phase rotation do not matter, we consider the squeezed-vacuum state with the covariance matrix $\sigma_r = {\rm diag}(r,1/r)$, where the squeezing parameter is between $r=0$ for infinitely $x$-quadrature squeezed state and $r=1$ for the vacuum state, as potentially optimal pure single-mode Gaussian probe state. Note that $p$-quadrature squeezed states can be obtained by a phase rotation from the $x$-quadrature squeezed states, so without loss of generality we assume  $0\leqslant r\leqslant1$. 
By using Equations~(\ref{eq:CM-time}) and (\ref{eq:QFI-Gaussian}), we can find an analytical expression for the QFI rate
	\begin{align}\label{eq:QFIRate-single}
		\hspace{-1mm}\tilde{\cal F}^{\rm Q}(\sigma_r;t) = \frac{(1-e^{-\gamma  t})^2(\partial_T \nu)^2\left([\sigma_t]_{11}^2+[\sigma_t]_{22}^2 + 2\right)}{2t([\sigma_t]_{11}^2[\sigma_t]_{22}^2 - 1)}. 
	\end{align}
	In Figure.~\ref{fig:QFI_rates_single} we depict the (rate of the) QFI against time for the vacuum state and for a squeezed state. It appears that the QFI rate is optimal when we perform the measurement as quickly as possible. 
	In the limit of short times and high squeezing ($r\ll 1$) and by keeping the most relevant terms in $t$ and $r$, we have
	\begin{align}\label{eq:QFI_small_single}
		\tilde{\cal F}^{\rm Q}(\sigma_{r\ll1} ; \gamma t\ll 1) = \frac{\gamma (\partial_T\nu)^2}{2(\gamma t \nu^2 + 2\nu r)}.
	\end{align}
	It is clear from this expression that the optimal measurement time approaches zero. Also, note that the rate can get arbitrarily large if we choose $\gamma t$ and $r$ sufficiently small. This implies that nonclassicality in terms of squeezing can significantly enhance precision.
    While the behavior of Fig.~\ref{fig:QFI_rates_single} depends on the temperature and the frequency (in particular the ratio $\omega/T$ is determinant through the parameter $\nu=\coth(\omega/2T)$) our numerics {in Appendix B show that that nonclassicality in the form of squeezing improves the QFI rate for other temperatures as well.}
 
	%
	Let us now explore the role of entanglement. To this aim, we assume possession of a secondary mode that does not undergo the dissipative dynamics but is rather initially entangled with the probe system---see Fig.~\ref{fig:set_up}. The dynamics can be simply extended to include the second mode by taking $X_t\to X_t\oplus I_2$, and $Y_t\to Y_t\oplus 0_2$,
	where $0_2$ is the $2\times 2$ null matrix. Equation (\ref{eq:CM-time}) is generalized as
	\begin{align}
		\sigma_t =  \left(\begin{array}{cc}
			e^{-\gamma t} A + (1-e^{-\gamma t})\nu I_2 & e^{-\frac{\gamma t}{2}} C \\
			e^{-\frac{\gamma t}{2}} C^T & B
		\end{array}\right),
	\end{align}
	where $A$, $B$ and $C$ are $2\times2$ matrices making up the initial covariance matrix at $t=0$. As we can see, at long times the correlations vanish. However, one can harness the correlations for enhanced thermometry at short evolution times. As the initial input, we choose the two-mode squeezed vacuum state given by
	\begin{align}\label{eq:2MSV}
		\sigma_{2,r} \coloneqq \left( \begin{array}{ccc}
			r^{-1}  \, {I} & \sqrt{r^{-2}-1}  \, {Z} \\
			\sqrt{r^{-2}-1}  \, {Z}  & r^{-1}   \, {I}
		\end{array} 
		\right),
	\end{align}
	with $0<r \leqslant 1$ being the squeezing parameter and $Z={\rm diag}(1,-1)$. This covariance matrix represents a pure entangled state iff $r <1$. 
	Notice that this scenario is at least as good as the one with a single mode squeezed state with the same squeezing parameter $\sigma_{r}$. If one performs a local Gaussian measurement described by $\sigma^{\rm M}$ on the auxiliary mode of (\ref{eq:2MSV}), the probe state collapses into Gaussian states with the covariance matrix $\sigma_{{\rm PS}} = r^{-1}  I - (r^{-2}-1) Z(r^{-1} I + \sigma^{\rm M})^{-1} Z$. Setting $\sigma^{\rm M} =\lim_{s\to 0} {\rm diag}(s,1/s)$ (homodyne), the covariance matrix of the probe states becomes $\sigma_{{\rm PS}} = \sigma_{r}$, that is a single-mode squeezed state. However, by performing joint measurements more information may be extracted.
	
\begin{figure}
		\centering
		\includegraphics[width=.3\columnwidth]{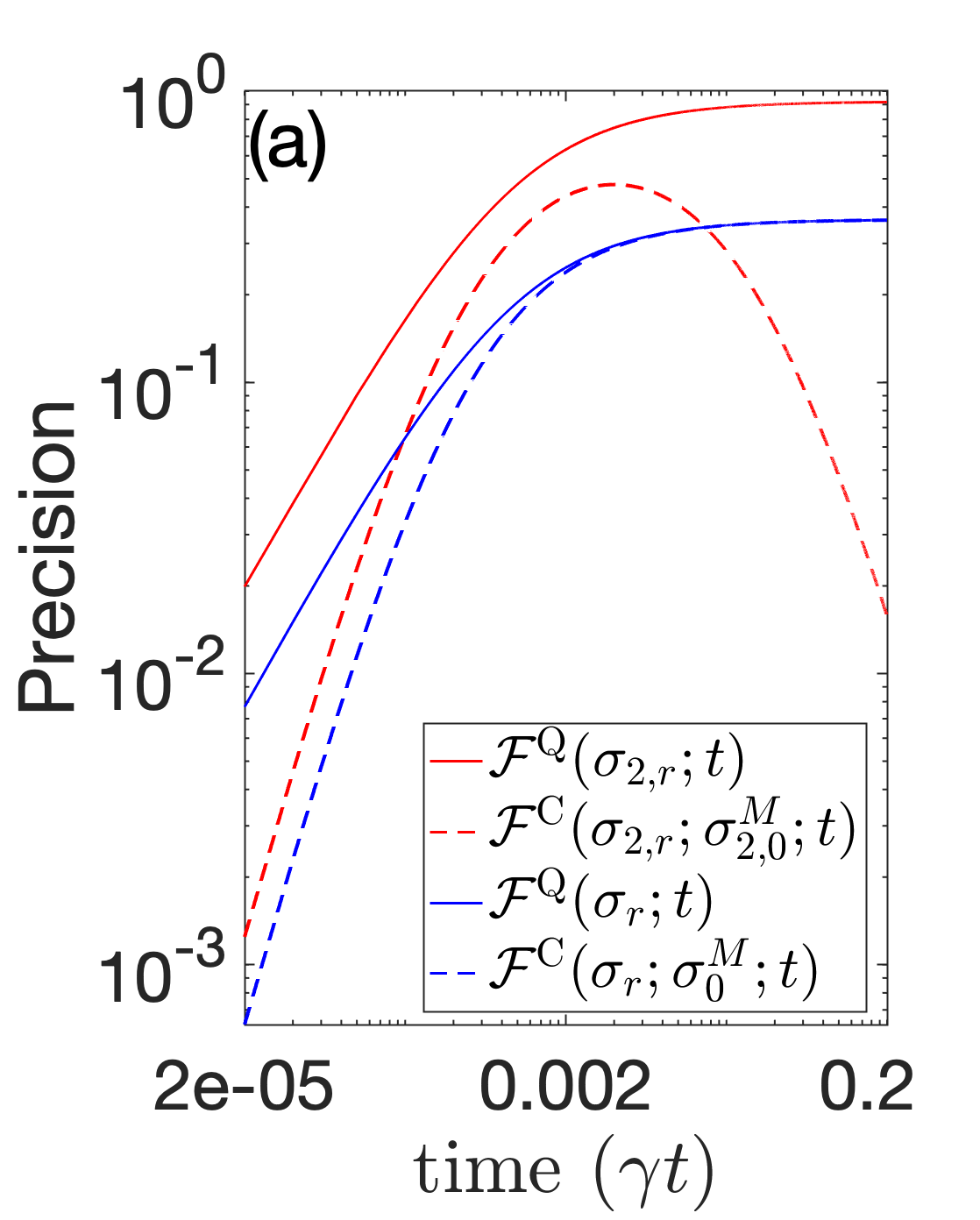}
		\includegraphics[width=.3\columnwidth]{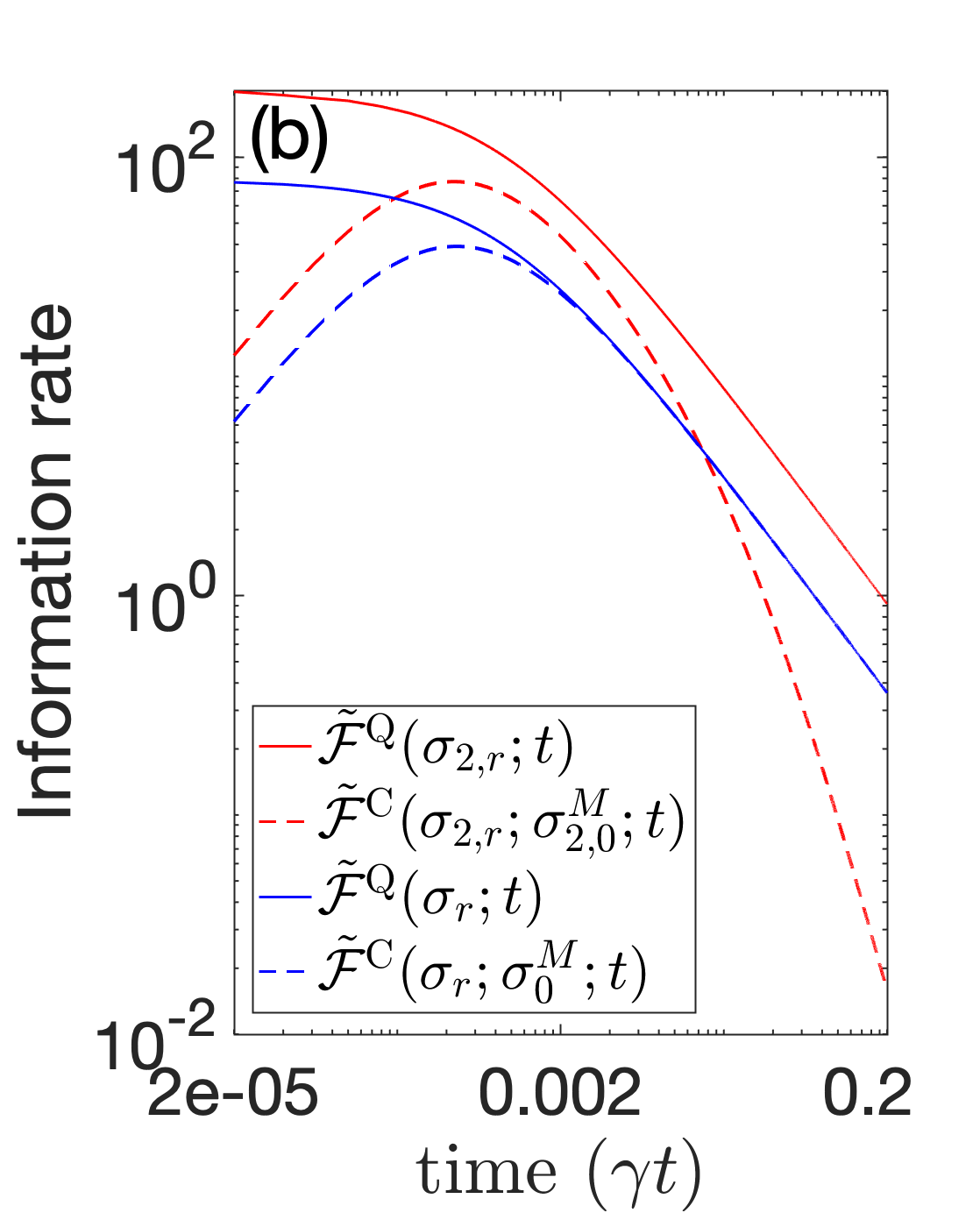}
		\caption{Same as Fig.~\ref{fig:QFI_rates_single}, but comparing the single-mode squeezed states with the two-mode squeezed states for the probe. By using entanglement the QFI saturates to its steady state value faster---at $\gamma t \gg 1$ the solid red curve will catch up with the solid blue one---making them even better than single mode squeezed states. Here, we examine the performance of a specific joint Gaussian measurement, consisting of a $50:50$ beamsplitter followed by two homodyne measurements.
			Evidently, even this simple Gaussian measurement can exploit the improvement offered by 
			the entangled Gaussian state, making them comparable to the single mode squeezing interrogated with optimal nonGaussian measurement. The rest of the parameters are taken similarly to Fig.~\ref{fig:QFI_rates_single}. $T=1$, $\gamma=0.2$, $r=10^{-3}$, $\omega=1$.}
		\label{fig:QFI_rates_two_modes}
	\end{figure}

 By using Eq.~(\ref{eq:QFI-Gaussian}), the QFI rate for the two-mode probe can be calculated and compared with the result of the single-mode probe. Our simulations depicted in Fig.~\ref{fig:QFI_rates_two_modes} confirm that indeed at short times the QFI rate can gain up to a two-fold improvement over the single mode squeezed state. Note, however, that this improvement requires joint measurements, which may not be practical in general. In the following, we show that by restricting to Gaussian measurements, which are readily available, one can still exploit the nonclassicality in probe states for thermometry precision.
	
\subsection{Result (iii): Optimality of Gaussian measurements}We now restrict to a fully Gaussian scenario where the measurements are Gaussian as well. In Appendix~\ref{app:opt_measure}, we analytically find the optimal Gaussian measurement for single-mode Gaussian probe states,  extending some of the results of~\cite{cenni2021thermometry} to dynamical scenarios. We prove that the optimal Gaussian measurement has a diagonal covariance matrix $\sigma^{\rm M}_L = {\rm diag}(L,1/L)$, which becomes homodyne for short times. By using  Eqs.~(\ref{eq:CM-time}) and (\ref{eq:CFI-Gaussian}) for the single-mode squeezed vacuum probe state with $\sigma_{r}$ and homodyne measurement on the $x$-quadrature ($L=0$), the  CFI rate is given by
	\begin{align}\label{eq:CFIRate-sing-hom}
		\tilde{\cal F}^{\rm C}(\sigma_{r};\sigma^{\rm M}_0;t)=\frac{(1-e^{-\gamma  t})^2({\partial_T \nu})^2}{2t(re^{-\gamma  t} + (1-e^{-\gamma  t})\nu)^2}.
	\end{align}
	We can check that $\tilde{\cal F}^{\rm C}(\sigma_{r\ll1};\sigma^{\rm M}_0;t\ll1) \approx \gamma^2 t(\partial_T\nu)^2/[2(r+\gamma t \nu)^2]$. By comparing this with \eqref{eq:QFI_small_single} and for $r\to 0$ one sees that $\tilde{\cal F}^{\rm Q}(\sigma_{r\ll 1};t\ll 1)\approx\tilde{\cal F}^{\rm C}(\sigma_{r\ll 1};\sigma^{\rm M}_0;t\ll 1)\approx (\partial_T\nu )^2/(2 t \nu^2)$ which suggests that homodyne measurement is the optimal measurement in the regime of large squeezing and at short times. For a finite squeezing parameter, unlike the QFI rate, the best interrogation time is not zero and roughly satisfies $t^* = {r}/(\gamma\nu)$, which decreases by increasing the squeezing and the dissipation rate $\gamma$. We have $\tilde{\cal F}^{\rm C}(\sigma_{r};\sigma^{\rm M}_0;t^*)\approx \gamma(\partial_T\nu )^2/(8 r \nu)$.

	Let us now discuss the role of entanglement in two-mode squeezed vacuum states using joint Gaussian measurements. In general, joint Gaussian measurements can be modeled in terms of two local phase rotation operations, a beamsplitter and two single-mode Gaussian measurements~\cite{JointGM}. However, instead of optimizing over all joint Gaussian measurements, we consider a measurement that boosts the performance with respect to the single-mode scenario. Specifically, we take  $\sigma^{\rm M}_{2,0} \coloneqq S_{\rm BS} \sigma_{xp} S_{BS}^T$ where the probe mode after the interaction overlaps with the auxiliary mode on a balanced beamsplitter with the symplectic transformation $S_{\rm BS}$. Then at the output two homodyne measurements, one on $x$-quadrature and one on $p$-quadrature with the CM $\sigma_{xp}=\lim_{L\to 0}{\rm diag}(L,1/L,1/L,L)$, are performed.  In this case, the CFI rate reads (see Appendix~\ref{app:feasible_meas} for details)
	\begin{align}
		\tilde{\cal F}^{\rm C}(\sigma_{2,r};\sigma^{\rm M}_{2,0};t) = \frac{(\partial_T\lambda)^2}{t\lambda^2},
	\end{align}
	where $\lambda\coloneqq (1+e^{-\gamma t})r^{-1}+\nu(1-e^{-\gamma t})-2e^{\frac{-\gamma t}{2}}\sqrt{r^{-2}-1}$.
	As depicted in Fig.~\ref{fig:QFI_rates_two_modes} this fully Gaussian setting can exploit entanglement to overperform the best (nonclassical) single mode setting, including nonGaussian measurements. At short times and for small values of $r$, we have
	\begin{align}
		\tilde{\cal F}^{\rm C}(\sigma_{2,r};\sigma^{\rm M}_{2,0};t\ll 1) \approx \frac{16r^2\gamma^2 t(\partial_T\nu)^2}{[(\gamma t)^2 + 4r^2+4r\gamma t\nu]^2},
	\end{align}
	which shows the optimal time is $t^*\approx{r}/({\gamma\nu})$. We then have $\tilde{\cal F}^{\rm C}(\sigma_{2,r};\sigma^{\rm M}_{2,0};t^*)\approx 16 \gamma\nu^3 (\partial_T\nu)^2/(r(1+8\nu^2)^2)$, which roughly shows two times improvement in the CFI rate compared to the single-mode squeezed probe state and homodyne measurement.
 
\section{Discussion}\label{sec:Discussion}
 Our formalism for nonequilibrium thermometry of bosonic baths by means of continuous-variable systems can be used to describe many platforms such as impurities in cold gases, mechanical resonators, Josephson junctions, and quantum-optical systems. We set a bound on the thermometric performance of all classical states (i.e., those with non-negative $P$-function), based on showing that all coherent states are equally useful. We then showed that nonclassical squeezed-vacuum states can beat this bound. An interesting lesson is that, increasing the energy of probes in terms of the displacement in phase space (or the amplitude of oscillation) is not beneficial for thermometry. But rather the energy should be put into increasing the degree of squeezing. We have shown, in particular, that fully Gaussian scenarios, which can simply be implemented in the laboratory in terms of single-mode and two-mode squeezed vacuum probe states and homodyne measurements, are extremely useful in boosting thermometry precision when time is a resource.
	
Our results also complement recent developments in nonequilibrium thermometry~\cite{Pavel_nonequilibrium}, which describes a fully non-restricted scenario. There are three major differences between the two scenarios. (1) In~\cite{Pavel_nonequilibrium}, the probe system has a finite dimensional-Hilbert space and the optimal preparation is always the eigenstate of the Hamiltonian with the maximum energy. Extending this to the infinite-dimensional Hilbert space with unbounded operators would be non-trivial. However, in our method, we consider the Gaussian states of a continuous-variable probe system with an infinite-dimensional Hilbert space. Gaussian states can be easily prepared using linear optics, and we find the optimal probe state within the Gaussian family. An interesting future direction is to generalize the upper bound of \cite{Pavel_nonequilibrium} to continuous variable probes by considering a constraint on the average energy of the probe states. (2)	When there is no restriction on the measurement, it is best to perform the measurement as fast as possible. However, when restricting to the Gaussian measurements, there exists an optimal measurement time---which as we have shown scales with the initial squeezing in the state of the probe. This is because while the QFI for the optimal input state grows linearly with time, the CFI of Gaussian measurements grows quadratically with time (except in the unphysical case where the probe is initially infinitely squeezed and the CFI  scales linearly with time). (iii) unlike the findings of \cite{Pavel_nonequilibrium}, there is no Lamb-shift effect in our model since the Lamb-shift is temperature independent---See Appendix~\ref{app:Brownain}. We expect that our formalism finds applications in estimating the temperature in quantum optical platforms or within cold Bosonic gases where state-of-the-art techniques allow for the implementation of Gaussian measurements such as homodyne/heterodyne measurements or time-of-flight measurements.



	%
	
	\section*{Acknowledgements}We are thankful to P. Sekatski and M. Perarnau-Llobet for constructive discussions. MM was supported the DFG/FWF Research Unit FOR 2724 ‘Thermal machines in the quantum world’ and by  Swiss National Science Foundation NCCR SwissMAP.  SM is supported by a grant from Basic Sciences Research,
	Iran, Fund (No. BSRF-phys-399-05).
	\onecolumngrid
	\appendix	
 
\section{Case study: The quantum Brownian motion}\label{app:Brownain}
Consider a Harmonic oscillator---playing the role of a Brownian particle, namely an impurity atoms or a light mode---that is embedded inside a bosonic bath---representing, e.g., cold bosonic gases or radiation fields. The Brownian particle is used as our probe to take the temperature of the bath. The total Hamiltonian reads
\begin{align}
    H = H_p + H_B + H_I.
    \label{eq:total_H}
\end{align}
Here, $H_p$ is the free Hamiltonian of the oscillator
\begin{align}
    H_p & = \omega_0 a^{\dagger} a,
\end{align}
with $a$ and $a^{\dagger}$ being the Bosonic creation and annihilation operators and $\omega_0$ is the bare frequency of the probe. 
Moreover, $H_B$ is the Hamiltonian of the bath given by
\begin{align}\label{eq:Bath_Ham_Brown}
    H_{B} = \sum_{\mu} \omega_{\mu} b^{\dagger}_{\mu}b_{\mu},
\end{align}
with $a_{\mu}$, $b^{\dagger}_{\mu}$, and $\omega_{\mu}$ being the creation, annihilation,and the bare frequency of the $\mu$th oscillator in the bath. Note that the creation and annihilation operators satisfy the commutation relation $[a,a^{\dagger}] = 1$, $[b_{\mu},b^{\dagger}_{\nu}] = \delta_{\mu,\nu}$.
Finally, $H_I$ is the linear interaction between the probe and the bath that we assume to read
\begin{align}
    H_I &=a^{\dagger}\otimes\sum _{\mu} g_{\mu} b_\mu +\text{ h.c.},  
\end{align}
where $g_{\mu}$ is the coupling strength and h.c. refers to Hermitian conjugate. Note that by defining the quadrature operators in the usual way, one can rewrite the interaction Hamiltonian in terms of Hermitian canonical operators
\begin{equation}\label{eq:Hermitian_interaction}
    H_{I}=\sum_{k=1}^2 A_k\otimes B_k
\end{equation}
where
\begin{align*}
    A_{1}&=\frac{1}{\sqrt{2}}(a^{\dagger}+a)=x, 
    \qquad B_{1}=\frac{1}{\sqrt{2}}\sum _{\mu} g_{\mu} (b_{\mu}^{\dagger}+b_{\mu})=\sum _{\mu} g_{\mu}x_\mu,\\
\hspace{.2cm} A_{2}&=\frac{i}{\sqrt{2}}(a^{\dagger}-a)=p, \qquad B_{2}=\frac{i}{\sqrt{2}}\sum _{\mu} g_{\mu} (b^{\dagger}_{\mu} - b_{\mu})=\sum _{\mu} g_{\mu}p_\mu.
\end{align*}
In the limit of continuous bath modes, we will use the following general description for the spectral density
\begin{align}\label{eq:SD_def}
	J(\omega) = 
 2\pi \sum _{k} g^{2}_{k} \delta (\omega -\omega _{k}).
\end{align}
For now, we do not assume any specific shape for the spectral density and keep the analysis as general as possible.
As we prove below, the Gorini–Kossakowski–Sudarshan–Lindblad (GKLS) master equation for $\rho_p$ the state of probe mode reads
\begin{align}\label{eq:ME_Brownian}
    \frac{d\rho_p}{dt}  = -i\big[H_p~+\Delta H_p\,,\rho_p\big] + \Gamma_{\rm out}\bigg(\!a\,\rho_p\, a^{\dagger} - \frac{1}{2}\big\{a^{\dagger}a,\rho_p\big\}\!\bigg) + \Gamma_{\rm in}\bigg(\!a^{\dagger}\rho_p\,a - \frac{1}{2}\big\{aa^{\dagger},\rho_p\big\}\!\bigg)
    ,
\end{align}
with $\Gamma_{\rm out} \coloneqq J(\omega_0)(N(\omega_0,T) + 1)$, $\Gamma_{\rm in} \coloneqq J(\omega_0)N(\omega_0,T)$, and $N(\omega_0,T) = [\exp(\omega_0/T) - 1]^{-1}$. The term $\Delta H_P$ represents a [Lamb] shift, which as we show below is temperature independent and thus the coherent part has no impact on the thermometry performance.
In the following subsection, we derive the master equation governing the damped harmonic oscillator dynamics in two ways, and particularly prove that the shift term is independent of temperature.

\subsection{Derivation of the GKLS master equation \eqref{eq:ME_Brownian}}
In the first approach, one can start from the von-Neumann equation ~\cite{rivas2011open, breuer_theory_2002}. 
We start by writing the evolution of the total system in the interaction picture
\begin{align}
    \dfrac{d \bm{\rho} (t)}{dt} =-i \big[ \bm{H}_{I} (t), \bm{\rho} (t) \big] ,
    \label{eq-vn}
\end{align}
where we use boldface letters for the operators in the interaction picture
\begin{align}
    \bm{\rho} (t) &=U^{\dagger}_{p} (t,0) \otimes U^{\dagger}_{B} (t,0) ~ \rho(t) ~ U_{p} (t,0) \otimes U_{B} (t,0) , 
    \label{eq-z1} \\
    \bm{H}_{I} (t) &=U^{\dagger}_{p} (t,0) \otimes U^{\dagger}_{B} (t,0) ~ H_{I} ~ U_{p} (t,0) \otimes U_{B} (t,0) .
    \label{eq-z2}
\end{align}
with $U_{p} (t,0) = e^{-itH_{p}}$ and $U_{B} (t,0) = e^{-itH_{B}}$, and $\rho(t)$ is 
the joint state of the probe and bath systems.
By integrating Eq. \eqref{eq-vn} we obtain
\begin{align}
     \bm{\rho} (t) = \bm{\rho} (0) -i\int_0^t ds \big[ \bm{H}_{I} (s), \bm{\rho} (s) \big],
\end{align}
which by recursive replacement reads
\begin{align}
    \bm{\rho} (t) = \bm{\rho} (0) -i\int_0^t ds \big[\bm{H}_{I} (s), \bm{\rho} (0)\big] -\int_0^t ds \int_0^s dr \big[ \bm{H}_{I} (s),\left[ \bm{H}_{I} (r), \bm{\rho} (r) \right]\!\big].
\end{align}
If we differentiate again, we have
\begin{align}
    \dfrac{d \bm{\rho} (t)}{dt} = -i \big[ \bm{H}_{I} (t), \bm{\rho} (0) \big] - \int ^{t} _{0} ds \big[ \bm{H}_{I} (t), \left[ \bm{H} _{I}(s), \bm{\rho}(s) \right]\! \big] .
    \label{eq-z3}
\end{align}
Now, by taking partial trace over the degrees of freedom of bath we have
\begin{align}
    \dfrac{d \bm{\rho} _{p} (t)}{dt} = -i\, \mathrm{Tr}_{B}\!\big[ \bm{H}_{I} (t), \bm{\rho} (0) \big] - \int ^{t} _{0} ds\, \mathrm{Tr}_{B}\!\big[ \bm{H}_{I} (t), \left[ \bm{H} _{I}(s), \bm{\rho}(s) \right]\!\big] .
    \label{eq-z4}
\end{align}
Considering that the initial state of the total system is a product state $\bm{\rho} (0) =\bm{\rho} _{p} (0) \otimes \bm{\rho} _{B}$, and that the bath is initially in a state which commutes with $H_{B}$ such as the thermal state, i.e. $\rho_{B}=e^{-H_{B}/T}/\mathrm{Tr}_{B}(e^{-H_{B}/T})$, the first term of Eq. \eqref{eq-z4} vanishes ~\cite{breuer_theory_2002}.
In this step, we employ the Born approximation, where the coupling between the probe and the bath is considered weak such that the effect of the system on the bath can be neglected; this implies that for the sufficiently large bath one can assume the bath state is time-independent and the probe-bath remain uncorrelated at all times, $\bm{\rho} (s) \approx \bm{\rho} _{p} (s) \otimes \bm{\rho} _{B}$ ~\cite{breuer_theory_2002,lidar2020lecture}.
Thus, Eq. \eqref{eq-z4} becomes
\begin{align}
    \dfrac{d \bm{\rho} _{p} (t)}{dt} & = - \int ^{t} _{0} ds\, \mathrm{Tr}_{B}\!\big[ \bm{H}_{I} (t), \left[ \bm{H} _{I}(s), \bm{\rho}_{p}(s) \otimes \bm{\rho} _{B} \right]\! \big]\nonumber\\
    & = - \int ^{t} _{0} ds\, \mathrm{Tr}_{B}\! \Big[ \bm{H}_{I}(t) \bm{H}_{I}(t-s) \bm{\rho}_{p}(t-s)\otimes \bm{\rho}_{B} - \bm{H}_{I}(t-s)\bm{\rho}_{p}(t-s)\otimes \bm{\rho}_{B} \bm{H}_{I}(t) +\text{h.c.} \Big] \nonumber\\
    & = -\sum_{kk^{\prime}}\int_0^t ds \Big(\! {\bm{A}}_k(t){\bm{A}}_{k^{\prime}}(t-s)\bm{\rho}_p(t-s)\! -\! {\bm{A}}_{k^{\prime}}(t-s) \bm{\rho}_p(t-s) {\bm{A}}_k(t) \!\Big) \mathrm{Tr}_{B}\!\big[{\bm{B}}_{k}(t){{\bm{B}}_{k^{\prime}}(t-s)\bm{\rho}_B}\big] + \text{h.c.} ,\nonumber\\
    & = -\sum_{kk^{\prime}}\int_0^t ds \Big(\! {\bm{A}}_k(t){\bm{A}}_{k^{\prime}}(t-s)\bm{\rho}_p(t)\! -\! {\bm{A}}_{k^{\prime}}(t-s) \bm{\rho}_p(t) {\bm{A}}_k(t)\! \Big) \mathrm{Tr}_{B}\!\big[{\bm{B}}_{k}(t){{\bm{B}}_{k^{\prime}}(t-s)\bm{\rho}_B}\big] + \text{h.c.} ,
    \label{eq-z5}
\end{align}
where we have applied $s \rightarrow t-s$ in the second line; in the third line we used the general form of the interaction Hamiltonian in the Schr\"{o}dinger picture as $H_{I}=\sum _{k}A_{k}\otimes B_{k}$, with $A_{k}$ and $B_{k}$ defined in Eq.~\eqref{eq:Hermitian_interaction}, and thus $\bm{A}_{k}(t)$ and $\bm{B}_{k}(t)$ are corresponding operators in the interaction picture
\begin{align}
\begin{split}
    \bm{A}_{1}(t)&=\frac{1}{\sqrt{2}}\Big(e^{i\omega_0 t}a^{\dagger} +  e^{-i\omega_0 t}a\Big),\qquad\bm{B}_{1}(t) =\dfrac{1}{\sqrt{2}}\sum _{\mu} g_{\mu} \Big( e^{it\omega _{\mu}} b_{\mu}^{\dagger} +e^{-it\omega _{\mu}} b_{\mu} \Big),\\
\bm{A}_{2}(t)&=\frac{i}{\sqrt{2}}\Big(e^{i\omega_0 t}a^{\dagger} -  e^{-i\omega_0 t}a\Big),\qquad\bm{B}_{2}(t)=\dfrac{i}{\sqrt{2}}\sum _{\mu} g_{\mu} \Big(e^{it\omega _{\mu}} b_{\mu}^{\dagger} - e^{-it\omega _{\mu}} b_{\mu} \Big);
\end{split}
    \label{eq-z7}
\end{align}
finally, in the last line, we use the Markov approximation by replacing $\bm{\rho}_p(t-s) \to \bm{\rho}_p(t)$.

To proceed further, we assume that $t\gg \tau_B$, where the bath correlation time scale $\tau_B$ is defined as the timescale beyond which the bath two-time correlation functions $\mathrm{Tr}_{B}[{{\bm{B}}_{k^{\prime}}(t-s){\bm{B}}_{k}(t)\bm{\rho}_B}]$ decay rapidly~\cite{lidar2020lecture, Albash_2012}. As a result, we can take the upper limit of the integral to infinity. We have 
\begin{align}
    \dfrac{d \bm{\rho} _{p} (t)}{dt} = \sum_{kk^{\prime}}\int_0^{\infty}\! ds \Big(\! {\bm{A}}_{k^{\prime}}(t-s) \bm{\rho}_p(t) {\bm{A}}_k(t)\! -\! {\bm{A}}_k(t){\bm{A}}_{k^{\prime}}(t-s)\bm{\rho}_p(t) \!\Big) \mathrm{Tr}_{B}\!\big[{\bm{B}}_{k}(t){{\bm{B}}_{k^{\prime}}(t-s)\bm{\rho}_B}\big] + \text{h.c.}.
    \label{eq-z6}
\end{align}

By using the Bosonic relations
\begin{align}
    \big\langle b^{\dagger}_{k} b_{l} \big\rangle & \equiv \mathrm{Tr}_{B}\!\big[\rho _{B} b^{\dagger}_{k} b_{l}\big]=\delta _{kl} N(\omega _{k} ,T) ,
    \label{eq-z8}\\
    \big\langle b_{k} b^{\dagger}_{l} \big\rangle &= \delta _{kl} \big(N(\omega _{k} ,T)+1\big) , 
    \label{eq-z9} \\
     \big\langle b^{\dagger}_{k} b^{\dagger}_{l} \big\rangle &=  \big\langle b_{k} b_{l} \big\rangle = 0 ,
    \label{eq-z10} 
\end{align}
where, $N(\omega _{k},T)$ is the average number of the $k$-th mode, we put together Eqs.~\eqref{eq-z7} and \eqref{eq-z6} and obtain
\begin{align}
    \dfrac{d \bm{\rho} _{p} (t)}{dt} = \sum_k g_k^2\int_0^{\infty} ds \Big(&a^{\dagger} \bm{\rho}_{p}(t) a N(\omega _{k},T) e^{-is(\omega _{0} -\omega _{k})}+ a \bm{\rho}_{p}(t) a^{\dagger}  \big( N(\omega _{k},T)+1 \big) e^{is(\omega _{0} -\omega _{k})}  
    \nonumber\\
     - & a^{\dagger}a \bm{\rho} _{p} (t)  \big( N(\omega _{k},T)+1 \big) e^{is(\omega _{0} -\omega _{k})} - a a^{\dagger} \bm{\rho}_{p}(t)  N(\omega _{k},T) e^{-is(\omega _{0} -\omega _{k})} \Big) + \text{h.c.}.
    \label{eq-z11} 
\end{align}
Now by using the following formula
\begin{align}
    \int _{0}^{\infty} ds e^{-is(\omega_0-\omega_k)} = \pi \delta (\omega_0-\omega_k) - i\mathbb{P}\dfrac{1}{\omega_0-\omega_k} ,
    \label{eq-z12} 
\end{align}
where $\mathbb{P}$ denotes the Cauchy principal value, and also 
using the spectral density function~\eqnref{eq:SD_def},
equation \eqref{eq-z11} simplifies to
\begin{align}
    \dfrac{d \bm{\rho} _{p} (t)}{dt} &= 
      \frac{1}{2} a^{\dagger} \bm{\rho}_{p}(t) a \left( \!J(\omega _{0}) N(\omega _{0} ,T) - \frac{i}{\pi}\mathbb{P} \int _{-\infty}^{\infty}\! d\omega\, \dfrac{J(\omega)N(\omega ,T)}{\omega_0 -\omega} \right)  \nonumber\\
    &~+ \frac{1}{2} a \bm{\rho}_{p}(t) a^{\dagger} \left(\! J(\omega _{0}) \left[N(\omega _{0} ,T)+1 \right] 
    + \frac{i}{\pi}\mathbb{P} \int _{-\infty}^{\infty} d\omega\, \dfrac{J(\omega) \left[N(\omega ,T)+1 \right] }{\omega _{0} -\omega} \right) \nonumber\\
    &~-\frac{1}{2} a^{\dagger}a \bm{\rho} _{p} (t) \left(\!J(\omega _{0}) \left[ N(\omega _{0} ,T)+1 \right] 
     + \frac{i}{\pi}\mathbb{P} \int _{-\infty}^{\infty}\! d\omega\, \dfrac{J(\omega) \left[N(\omega,T)+1 \right] }{\omega_0 -\omega} \right)  \nonumber\\
    &~- \frac{1}{2} aa^{\dagger} \bm{\rho}_{p}(t) \left(\! J(\omega _{0}) N(\omega _{0} ,T) 
     - \frac{i}{\pi} \mathbb{P} \int _{-\infty}^{\infty}\! d\omega\, \dfrac{J(\omega) N(\omega ,T)}{\omega _{0} -\omega} \right)    +\text{h.c.}~. 
    \label{eq-z14} 
\end{align}
Finally, the dynamics of the damped harmonic oscillator in the interaction picture reads
\begin{align}
\dfrac{d \bm{\rho} _{p} (t)}{dt} = & -i \Delta _{1}\big[ a^{\dagger}a , \bm{\rho} _{p} (t)  \big] + i \Delta _{2} \left[ aa^{\dagger} , \bm{\rho} _{p} (t)  \right] + J(\omega _{0}) \big(N(\omega _{0} ,T)+1 \big) \Big( a \bm{\rho} _{p} (t) a^{\dagger} -\dfrac{1}{2}\big\lbrace a^{\dagger} a ,\bm{\rho} _{p} (t) \big\rbrace\! \Big) \nonumber\\
&+J(\omega _{0}) N(\omega _{0} ,T) \Big( a ^{\dagger} \bm{\rho} _{p} (t) a -\dfrac{1}{2}\big\lbrace a a^{\dagger} ,\bm{\rho} _{p} (t) \big\rbrace\! \Big) ,
\label{eq-z15}
\end{align}
where
\begin{align}
\Delta _{1} &= \frac{1}{2\pi}\mathbb{P} \int _{-\infty}^{\infty}\! d\omega\, \dfrac{J(\omega) \left[ N(\omega ,T)+1 \right]}{\omega _{0} -\omega} ,
\label{eq-z16} \\
\Delta _{2} &= \frac{1}{2\pi}\mathbb{P} \int _{-\infty}^{\infty}\!  d\omega\, \dfrac{J(\omega) N(\omega ,T)}{\omega _{0} -\omega} .
\label{eq-z17} 
\end{align}
Since $[a,a^{\dagger}]=\mathbb{I}$, then:
\begin{align}
\dfrac{d \bm{\rho} _{p} (t)}{dt} = & -i \big[ \left(\Delta _{1} -\Delta _{2}\right) a^{\dagger}a\, , \bm{\rho} _{p} (t)  \big] +J(\omega _{0}) \big( N(\omega _{0} ,T)+1 \big) \Big( a \bm{\rho} _{p} (t) a^{\dagger} -\dfrac{1}{2} \big\lbrace a^{\dagger} a ,\bm{\rho} _{p} (t) \big\rbrace\! \Big) \nonumber\\
& +J(\omega _{0}) N( \omega _{0} ,T) \Big( a ^{\dagger} \bm{\rho} _{p} (t) a -\dfrac{1}{2} \big\lbrace a a^{\dagger} ,\bm{\rho} _{p} (t) \big\rbrace\!\Big) .
\label{eq-z18}
\end{align}
By using Eq. \eqref{eq-z1} the master equation in the Schr\"{o}dinger picture can also be obtained
\begin{align}
\dfrac{d \rho _{p} (t)}{dt}=-i \big[ H_{\rm LS}+\omega_{0} a^{\dagger}a , \rho _{p} (t) \big] +\mathcal{D} [\rho _{p} (t)] ,
\label{eq-z19}
\end{align}
where
\begin{align}
\mathcal{D}[\rho _{p}(t)]= J(\omega _{0}) \big(N(\omega _{0},T)+1 \big) \Big( a \rho _{p} (t) a^{\dagger} -\dfrac{1}{2}\big\lbrace a^{\dagger} a ,\rho _{p} (t) \big\rbrace\! \Big)+J(\omega _{0})N(\omega _{0} ,T) \Big( a ^{\dagger} \rho _{p} (t) a -\dfrac{1}{2}\big\lbrace a a^{\dagger} ,\rho _{p} (t) \big\rbrace\! \Big),
\label{eq-z21}
\end{align}
is the dissipator of the master equation.
and
\begin{align}
H_{\rm LS}=(\Delta _{1} -\Delta _{2}) a^{\dagger}a ,
\label{eq-z20}
\end{align}
is the Lamb-shift Hamiltonian which leads to a renormalization of bare particle Hamiltonian 
due to the particle-bath interaction. In general, it has been shown that the Lamb-shift terms have a significant impact on the dynamics ~\cite{PhysRevA.105.012208}, and even on thermometry~\cite{Pavel_nonequilibrium}. However, in our case, since
\begin{align}
\Delta _{1}-\Delta _{2}= \mathbb{P} \int _{-\infty}^{\infty} d\omega\dfrac{J(\omega)}{\omega _{0} -\omega} ,
\label{eq-z22}
\end{align}
it is clear that the Lamb-shift term is independent of $N(\omega_0, T)$ and therefore independent of the bath temperature. Thus, the presence of the Lamb shift does not affect the thermometry performance, in particular, the first term in Eqs.~\eqref{eq:QFI-Gaussian} and \eqref{eq:CFI-Gaussian} disappear and our no-go result holds. 

Note that a similar result is expected if the interaction Hamiltonian was considered to be of $H_I = (a+a^{\dagger})\sum_{\mu}g_{\mu}(b+b^{\dagger})$, which can be effectively reduced to our model by means of the secular approximation. See also Ref.~\cite{correa2023potential} for a thorough discussion on Lamb-Shift and its impact on the dynamics for the same model with an Ohmic spectral density with Lorentz-Drude cutoff.

In the second approach for deriving the master equation~\eqnref{eq:ME_Brownian}, one can use the standard closed form of the Lindblad equation to describe the time evolution of the damped harmonic oscillator ~\cite{breuer_theory_2002,lidar2020lecture}. For this purpose, we start from the following equation in the Schr\"{o}dinger picture
\begin{align}
\dfrac{d \rho _{p} (t)}{dt}=-i\big[H_{p}, \rho _{p}(t)\big]+\sum _{\omega} \sum _{k ,k^{\prime}} \Gamma _{k k^{\prime}}(\omega) \Big(\! A_{k^{\prime}}(\omega)\rho _{p}(t) A_{k}^{\dagger}(\omega)-A_{k}^{\dagger}(\omega)A_{k^{\prime}}(\omega)\rho _{p}(t)\! \Big) +\text{h.c.} ,
\label{eq-z23}
\end{align} 
where we assumed the interaction Hamiltonian~\eqref{eq:Hermitian_interaction} and define the coefficients $\Gamma _{k k^{\prime}}(\omega)$ and operators $A_{k}(\omega)$ in the following.
In the above equation, we introduce the one-sided Fourier transformation of the bath correlation functions as
\begin{align}
\Gamma _{k k^{\prime}}(\omega)=\int _{0} ^{\infty}\! ds\, e^{i\omega s} \left\langle {\bm B}_{k}(s){\bm B}_{k^{\prime}}(0) \right\rangle ,
\label{eq-z26}
\end{align}
where $\left\langle {\bm B}_{k}(s){\bm B}_{k^{\prime}}(0) \right\rangle = \mathrm{Tr}_{B}[{\bm B}_{k}(s){\bm B}_{k^{\prime}}(0) \rho _{B}]$ with ${\bm B}_{k}(t)$ being the bath operators in the interaction picture~\eqnref{eq-z7}. 
Substituting ${\bm B}_{1}(t)$ and ${\bm B}_{2}(t)$ into Eq. \eqref{eq-z26} and using Eqs. \eqref{eq-z8}-\eqref{eq-z10} we find
\begin{align}
&\Gamma _{11}(\omega)=\Gamma _{22}(\omega) =\sum _{k} \dfrac{g^{2}_{k}}{2} \bigg(\! N(\omega _{k}, T)\! \int _{0}^{\infty} ds e^{i(\omega +\omega _{k})s} +\big(N(\omega _{k}, T) +1\big) \!\int _{0}^{\infty} ds e^{i(\omega -\omega _{k})s}\! \bigg) ,
\label{eq-z32}\\
&\Gamma _{12}(\omega)=-\Gamma _{21}(\omega) =\sum _{k} -i\dfrac{g^{2}_{k}}{2} \bigg(\! N(\omega _{k}, T)\! \int _{0}^{\infty} ds e^{i(\omega +\omega _{k})s} -\big(N(\omega _{k}, T) +1\big)\! \int _{0}^{\infty} ds e^{i(\omega -\omega _{k})s}\! \bigg) .
\label{eq-z33}
\end{align}
By considering a continuous spectrum for the bath and using spectral density function \eqref{eq:SD_def} and  Eq.~\eqref{eq-z12}, we get
\begin{align}
&\Gamma _{11}(\omega)=\Gamma _{22}(\omega) = \frac{1}{4} \big(N(\omega, T)+1\big) \big(J(\omega)-J(-\omega)\big) +\dfrac{i}{4\pi} \mathbb{P}\! \int _{-\infty}^{\infty}\! d\omega^{\prime} J(\omega^{\prime}) \bigg(\dfrac{N(\omega^{\prime}, T)}{\omega^{\prime} +\omega}+\dfrac{N(\omega^{\prime}, T)+1}{\omega^{\prime} -\omega} \bigg) ,
\label{eq-z34}\\
&\Gamma _{12}(\omega)=-\Gamma _{21}(\omega) =\frac{i}{4}\big(N(\omega, T)+1\big) \big(J(\omega)+J(-\omega)\big) +\dfrac{1}{4\pi} \mathbb{P}\! \int _{-\infty}^{\infty}\! d\omega^{\prime} J(\omega^{\prime}) \bigg(\dfrac{N(\omega^{\prime}, T)}{\omega^{\prime} +\omega}-\dfrac{N(\omega^{\prime}, T)+1}{\omega^{\prime} -\omega} \bigg) .
\label{eq-z35}
\end{align}

By using the eigenvectors and eigenvalues of the probe Hamiltonian, $H_{p}\vert n \rangle=(n+\dfrac{1}{2})\omega _{0} \vert n\rangle $, the operators $A_{k}(\omega)$ in~\eqref{eq-z23} are also defined as
\begin{align}
A_{k}(\omega)\coloneqq \sum _{(n'-n)\omega _{0}=\omega} \!\!\vert n\rangle\langle n \vert A_{k} \vert n'\rangle\langle n'\vert.
\label{eq-z25}
\end{align}
For $A_{1}=(a^{\dagger}+a) /\sqrt{2}$, we have
\begin{align}
A_{1}(\omega) &=\dfrac{1}{\sqrt{2}} \sum _{(n'-n)\omega _{0}=\omega} \!\!\vert n\rangle\langle n\vert (a^{\dagger}+a) \vert n'\rangle\langle n' \vert =\dfrac{1}{\sqrt{2}}\! \sum _{(n'-n)\omega _{0}=\omega}\!\!\! \Big(\! \sqrt{n'+1} \vert n\rangle\langle n'\vert \delta _{n, n'+1} + \sqrt{n'} \vert n\rangle\langle n'\vert \delta _{n, n'-1} \Big) ,
\label{eq-z36}
\end{align}
which leads to two Lindblad operators with corresponding frequencies $+\omega _{0}$ and $-\omega _{0}$ as
\begin{align}
A_{1}(+\omega _{0}) &=\dfrac{1}{\sqrt{2}} \sum _{n} \sqrt{n}\vert n-1\rangle\langle n\vert = \dfrac{1}{\sqrt{2}} a ,
\label{eq-z37}\\
A_{1}(-\omega _{0}) &=\dfrac{1}{\sqrt{2}} \sum _{n} \sqrt{n+1} \vert n+1\rangle\langle n\vert = \dfrac{1}{\sqrt{2}} a^{\dagger} 
\label{eq-z38} 
\end{align}
Similarly, for $A_2=i(a^{\dagger}-a) /\sqrt{2}$, we get other Lindblad operators 
\begin{align}
A_{2}(+\omega _{0}) &=-\dfrac{i}{\sqrt{2}} \sum _{n} \sqrt{n}\vert n-1\rangle\langle n\vert = -\dfrac{i}{\sqrt{2}} a,
\label{eq-z39}\\
A_{2}(-\omega _{0}) &=\dfrac{i}{\sqrt{2}} \sum _{n} \sqrt{n+1} \vert n+1\rangle\langle n\vert = \dfrac{i}{\sqrt{2}} a^{\dagger} .
\label{eq-z40} 
\end{align}
It is clear that for the damped harmonic oscillator we have $k,k^{\prime} \in \lbrace 1,2\rbrace$ and $\omega \in \lbrace \omega _{0}, -\omega _{0}\rbrace$, thus Eq. \eqref{eq-z23} becomes
\begin{align}
\dfrac{d \rho _{p} (t)}{dt}=&-i\left[H_{p}, \rho _{p}(t)\right]+\Gamma _{11}(\omega _{0}) \Big(\! A_{1}(\omega _{0}) \rho _{p}(t) A_{1} ^{\dagger}(\omega _{0})-A_{1} ^{\dagger}(\omega _{0})A_{1}(\omega _{0})\rho _{p}(t) \Big) \nonumber\\
&+ \Gamma _{12}(\omega _{0}) \Big(\! A_{2}(\omega _{0}) \rho _{p}(t) A_{1} ^{\dagger}(\omega _{0})-A_{1} ^{\dagger}(\omega _{0})A_{2}(\omega _{0})\rho _{p}(t) \Big) \nonumber\\
&+ \Gamma _{21}(\omega _{0}) \Big(\! A_{1}(\omega _{0}) \rho _{p}(t) A_{2} ^{\dagger}(\omega _{0})-A_{2} ^{\dagger}(\omega _{0})A_{1}(\omega _{0})\rho _{p}(t) \Big) \nonumber\\
&+ \Gamma _{22}(\omega _{0}) \Big(\! A_{2}(\omega _{0}) \rho _{p}(t) A_{2} ^{\dagger}(\omega _{0})-A_{2} ^{\dagger}(\omega _{0})A_{2}(\omega _{0})\rho _{p}(t) \Big) \nonumber\\
&+ \Gamma _{11}(-\omega _{0}) \Big(\! A_{1}(-\omega _{0}) \rho _{p}(t) A_{1} ^{\dagger}(-\omega _{0})-A_{1} ^{\dagger}(-\omega _{0})A_{1}(-\omega _{0})\rho _{p}(t) \Big) \nonumber\\
&+ \Gamma _{12}(-\omega _{0}) \Big(\! A_{2}(-\omega _{0}) \rho _{p}(t) A_{1} ^{\dagger}(-\omega _{0})-A_{1} ^{\dagger}(-\omega _{0})A_{2}(-\omega _{0})\rho _{p}(t) \Big) \nonumber\\
&+ \Gamma _{21}(-\omega _{0}) \Big(\! A_{1}(-\omega _{0}) \rho _{p}(t) A_{2} ^{\dagger}(-\omega _{0})-A_{2} ^{\dagger}(-\omega _{0})A_{1}(-\omega _{0})\rho _{p}(t) \Big) \nonumber\\
&+ \Gamma _{22}(-\omega _{0}) \Big(\! A_{2}(-\omega _{0}) \rho _{p}(t) A_{2} ^{\dagger}(-\omega _{0})-A_{2} ^{\dagger}(-\omega _{0})A_{2}(-\omega _{0})\rho _{p}(t) \Big) + \text{h.c.}.
\label{eq-z41}
\end{align}
By substituting Lindblad operators and considering $\Gamma _{k k^{\prime}}(\omega)=\gamma _{k k^{\prime}}(\omega)+i S_{k k^{\prime}}(\omega)$, Eq. \eqref{eq-z41} becomes
\begin{align}
&\dfrac{d \rho _{p} (t)}{dt}=-i\left[H_{p}, \rho _{p}(t)\right]\nonumber\\
&+\dfrac{1}{2}\big(\gamma _{11}(\omega _{0})+i S_{11}(\omega _{0})\big)\big(a \rho _{p}(t) a^{\dagger} - a^{\dagger} a\rho _{p}(t)\big)+\dfrac{1}{2}\big(\gamma _{11}(\omega _{0})-i S_{11}(\omega _{0})\big)\big(a \rho _{p}(t) a^{\dagger} -\rho _{p}(t) a^{\dagger} a\big) \nonumber\\
&-\dfrac{i}{2}\big(\gamma _{12}(\omega _{0})+i S_{12}(\omega _{0})\big)\big(a \rho _{p}(t) a^{\dagger} - a^{\dagger} a\rho _{p}(t)\big)+\dfrac{i}{2}\big(\gamma _{12}(\omega _{0})-i S_{12}(\omega _{0})\big)\big(a \rho _{p}(t) a^{\dagger} -\rho _{p}(t) a^{\dagger} a\big]) \nonumber\\
&-\dfrac{i}{2}\big(\gamma _{12}(\omega _{0})+i S_{12}(\omega _{0})\big)\big(a \rho _{p}(t) a^{\dagger} - a^{\dagger} a\rho _{p}(t)\big)+\dfrac{i}{2}\big(\gamma _{12}(\omega _{0})-i S_{12}(\omega _{0})\big)\big(a \rho _{p}(t) a^{\dagger} -\rho _{p}(t) a^{\dagger} a\big) \nonumber\\
&+\dfrac{1}{2}\big(\gamma _{11}(\omega _{0})+i S_{11}(\omega _{0})\big)\big(a \rho _{p}(t) a^{\dagger} - a^{\dagger} a\rho _{p}(t)\big)+\dfrac{1}{2}\big(\gamma _{11}(\omega _{0})-i S_{11}(\omega _{0})\big)\big(a \rho _{p}(t) a^{\dagger} -\rho _{p}(t) a^{\dagger} a\big) \nonumber\\
&+\dfrac{1}{2}\big(\gamma _{11}(-\omega _{0})+i S_{11}(-\omega _{0})\big)\big(a^{\dagger} \rho _{p}(t) a -a a^{\dagger} \rho _{p}(t)\big)+\dfrac{1}{2}\big(\gamma _{11}(-\omega _{0})-i S_{11}(-\omega _{0})\big)\big(a^{\dagger} \rho _{p}(t) a -\rho _{p}(t)a a^{\dagger} \big) \nonumber\\
&+\dfrac{i}{2}\big(\gamma _{12}(-\omega _{0})+i S_{12}(-\omega _{0})\big)\big(a^{\dagger} \rho _{p}(t) a -a a^{\dagger} \rho _{p}(t)\big)-\dfrac{i}{2}\big(\gamma _{12}(-\omega _{0})-i S_{12}(-\omega _{0})\big)\big(a^{\dagger} \rho _{p}(t) a -\rho _{p}(t) a a^{\dagger} \big) \nonumber\\
&+\dfrac{i}{2}\big(\gamma _{12}(-\omega _{0})+i S_{12}(-\omega _{0})\big)\big(a^{\dagger} \rho _{p}(t) a -a a^{\dagger} \rho _{p}(t)\big)-\dfrac{i}{2}\big(\gamma _{12}(-\omega _{0})-i S_{12}(-\omega _{0})\big)\big(a^{\dagger} \rho _{p}(t) a -\rho _{p}(t)a a^{\dagger} \big) \nonumber\\
&+\dfrac{1}{2}\big(\gamma _{11}(-\omega _{0})+i S_{11}(-\omega _{0})\big)\big(a ^{\dagger}\rho_{p}(t) a- a a^{\dagger} \rho _{p}(t)\big)+\dfrac{1}{2}\big(\gamma _{11}(-\omega _{0})-i S_{11}(-\omega _{0})\big)\big(a^{\dagger} \rho_{p}(t) a -\rho _{p}(t)a a^{\dagger} \big) .
\label{eq-z42}
\end{align}
The above equation then simplifies to
\begin{align}
&\dfrac{d \rho _{p} (t)}{dt}=-i\left[H_{p}\,, \rho _{p}(t)\right]\nonumber\\
&+2\big(\gamma _{11}(\omega _{0})+ S_{12}(\omega _{0})\big) \Big( a\rho_{p}(t) a^{\dagger} -\dfrac{1}{2}\big\lbrace a^{\dagger}a ,\rho _{p}(t) \big\rbrace\! \Big) -i\big(S _{11}(\omega _{0})- \gamma _{12}(\omega _{0})\big) \big[ a^{\dagger}a ,\rho _{p}(t) \big] \nonumber\\
&+2\big(\gamma _{11}(-\omega _{0})- S_{12}(-\omega _{0})\big) \Big( a^{\dagger}\rho_{p}(t) a -\dfrac{1}{2}\big\lbrace aa^{\dagger} ,\rho _{p}(t) \big\rbrace \!\Big) -i\big(S _{11}(-\omega _{0})+ \gamma _{12}(-\omega _{0})\big) \big[ aa^{\dagger} ,\rho _{p}(t) \big].
\label{eq-z43}
\end{align}
The coefficients in the above equation are easily calculated to be 
\begin{align}
&\gamma _{11}(\omega _{0})+ S_{12}(\omega _{0})=\frac{1}{2}J(\omega_{0}) \big( N(\omega_{0}, T)+1 \big) ,
\label{eq-z44}\\
&S_{11}(\omega _{0})- \gamma _{12}(\omega _{0})=\Delta_{1},
\label{eq-z45}\\
&\gamma _{11}(-\omega _{0})- S_{12}(-\omega _{0})=-\frac{1}{2}J(\omega_{0}) \big( N(-\omega_{0}, T)+1 \big) = \frac{1}{2}J(\omega_{0})N(\omega_{0},T),
\label{eq-z46}\\
&S _{11}(-\omega _{0})+\gamma _{12}(-\omega _{0})=-\Delta_{2},
\label{eq-z47}
\end{align}
where we have used $N(-\omega_{0},T)=-N(\omega_{0},T)-1$. Finally by using $[a,a^{\dagger}]=\mathbb{I}$, the quantum master equation for the damped harmonic oscillator in the Schr\"{o}dinger picture is obtained
\begin{align}
\dfrac{d \rho _{p} (t)}{dt}=&-i \left[ (\Delta _{1}-\Delta _{2}+\omega _{0}) a^{\dagger}a , \rho _{p} (t) \right] + J(\omega _{0}) \big(N(\omega _{0} ,T)+1 \big) \Big( a \rho _{p} (t) a^{\dagger} -\dfrac{1}{2} \big\lbrace a^{\dagger} a ,\rho _{p} (t) \big\rbrace \!\Big) \nonumber\\
& +J(\omega _{0})N(\omega _{0} ,T) \Big( a ^{\dagger} \rho _{p} (t) a -\dfrac{1}{2} \big\lbrace a a^{\dagger} ,\rho _{p} (t) \big\rbrace \!\Big),
\label{eq-z48}
\end{align}
which has the same form as in Eq. \eqref{eq-z19}.
  

\subsection{The master equation for the moments}
To write down the master equation for the first and the second moments we need to bring the master equation in the standard form \cite{PhysRevA.94.052129}.
We can define new jump operators to absorb the decay rates, i.e.,
\begin{align}
    L^{+}& =\sqrt{\Gamma_{\rm in}}a = \sqrt{\frac{\Gamma_{\rm in}}{2}}(1,~i) R\eqqcolon (c^+)^T R
,\\ 
L^{-}&=\sqrt{\Gamma_{\rm out}}a^{\dagger} = \sqrt{\frac{\Gamma_{\rm out}}{2}}(1,~-i) R \eqqcolon (c^-)^T R,
\end{align}
where the decay rates are defined as $\Gamma_{\rm in} \coloneqq J(\omega_0)N(\omega_0,T)$, and $\Gamma_{\rm out} \coloneqq J(\omega_0)[N(\omega_0,T)+1]$, and $R=(x,~p)^T$ is the vector of the canonical operators.
Then, the Lindbladian master equation~\eqref{eq:ME_Brownian} can be equivalently written for the displacement vector and covariance matrix:
\begin{align}
\label{ME_Cov}
    \frac{d}{dt}d &= A~d,\nonumber\\
    \frac{d}{dt}\sigma &= A~\sigma~+~\sigma~A^T~+~D,
\end{align}
with $A = \Omega~[G - {\rm Im}(CC^{\dagger})]$, $D = -2\Omega~{\rm Re}(CC^{\dagger})~\Omega$, and $C = \left(c^{+}, c^{-}\right)$.
Substituting for $C$, we find that
\begin{align}
    {\rm Re}~CC^{\dagger} &=\frac{J(\omega_0)}{2}\big(2N(\omega_0,T)+1\big)\!\left(
    \begin{array}{cc}
        1 &  0\\
        0 & 1
    \end{array}
    \right)
    =  \frac{J(\omega_0)}{2}\coth\!\Big(\frac{\omega_0}{2T}\Big)
    I_2
    \\
    {\rm Im}~CC^{\dagger}
    & = \frac{J(\omega_0)}{2}\left( \begin{array}{cc}
    	0 & -1\\
    	1 & 0
    \end{array}
    \right).
\end{align}
Putting everything together, we have
\begin{align}
	D = J(\omega_0)\coth\!\Big(\frac{\omega_0}{2T}\Big)I_2,
\end{align}
and
\begin{align}
    A = \left(
    \begin{array}{cc}
        -\frac{J(\omega_0)}{2} & \omega_0 \\
        -\omega_0 & -\frac{J(\omega_0)}{2}
    \end{array}
    \right).
\end{align}
The time dependent solution to the ``Lyapunov differential equation" \eqref{ME_Cov} is given by
\begin{align}\label{eq:Lyapunov_time_solution}
    d(t) & = e^{t A} d,\nonumber\\
	\sigma(t) & = e^{t A}\sigma(0)e^{tA^{T}} + \int_0^t ds e^{(t-s)A} D e^{(t-s)A^{T}}.
\end{align}
Note that since the real part of the eigenvalues of $A$ are always negative, we will have a steady state.
One can find the exponential of $A$, e.g., by using its Jordan form. It reads 
\begin{align}
    {\rm e}^{tA} & = {\rm e}^{-J(\omega_0) t}~\left(
    \begin{array}{cc}
        \cos(\omega_0~t) & \sin(\omega_0~t) \\
        -~\sin(\omega_0~t) & \cos(\omega_0~t)
    \end{array}
    \right)\eqqcolon e^{-J(\omega_0)t/2}O_t,
\end{align}
and thus
\begin{align}\label{eq:CV_J}
    d(t) & = e^{-J(\omega_0)t/2}O_t d,\nonumber\\
    \sigma(t) & = e^{-J(\omega_0)t}O_t\sigma O^T_t + \big(1-e^{-J(\omega_0)t}\big)\nu I_2.
\end{align}
with $\nu=\coth{({\omega_0}/{2T})}$. Again, note that the rotation $O_t$ is temperature independent and has no thermometric value; thus we work in the interaction picture. The steady state moments can be easily found to be the thermal ones, i.e., $d( \infty) = 0$, and  $\sigma(\infty) = \sigma_T = \coth{({\omega_0}/{2T})}I_2$.
One sees that the speed of converging to the thermal state is given by $J(\omega_0)$. That is, the spectral density at the frequency of the oscillator plays the role of the decay rate $\gamma$ in Eq.~\eqref{eq:CM-time}. 
\section{More simulations for the QFI rate}\label{app:more_simulations}
\begin{figure}[h!]
		\centering
		\includegraphics[width=.3\linewidth]{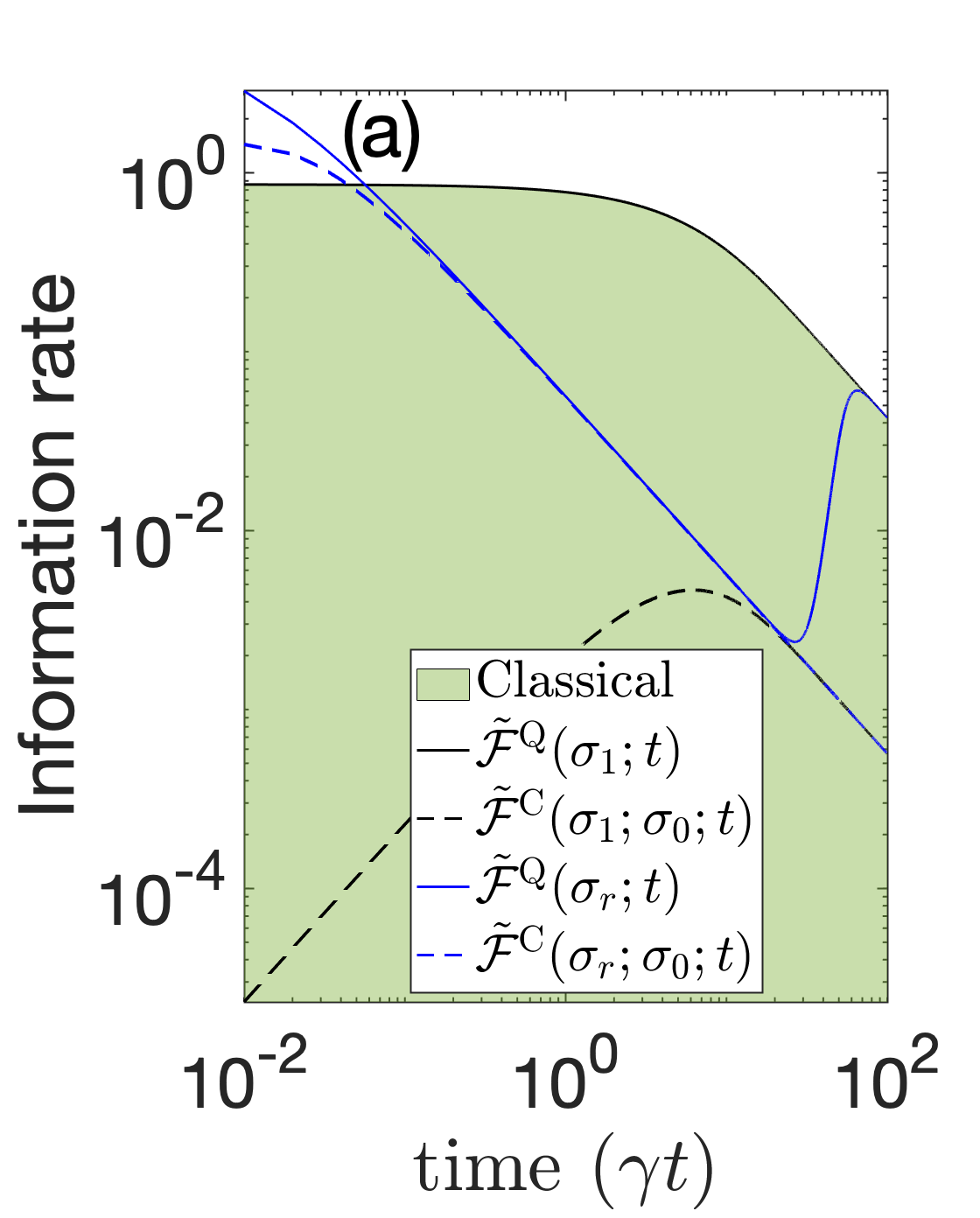}
		\includegraphics[width=.3\linewidth]{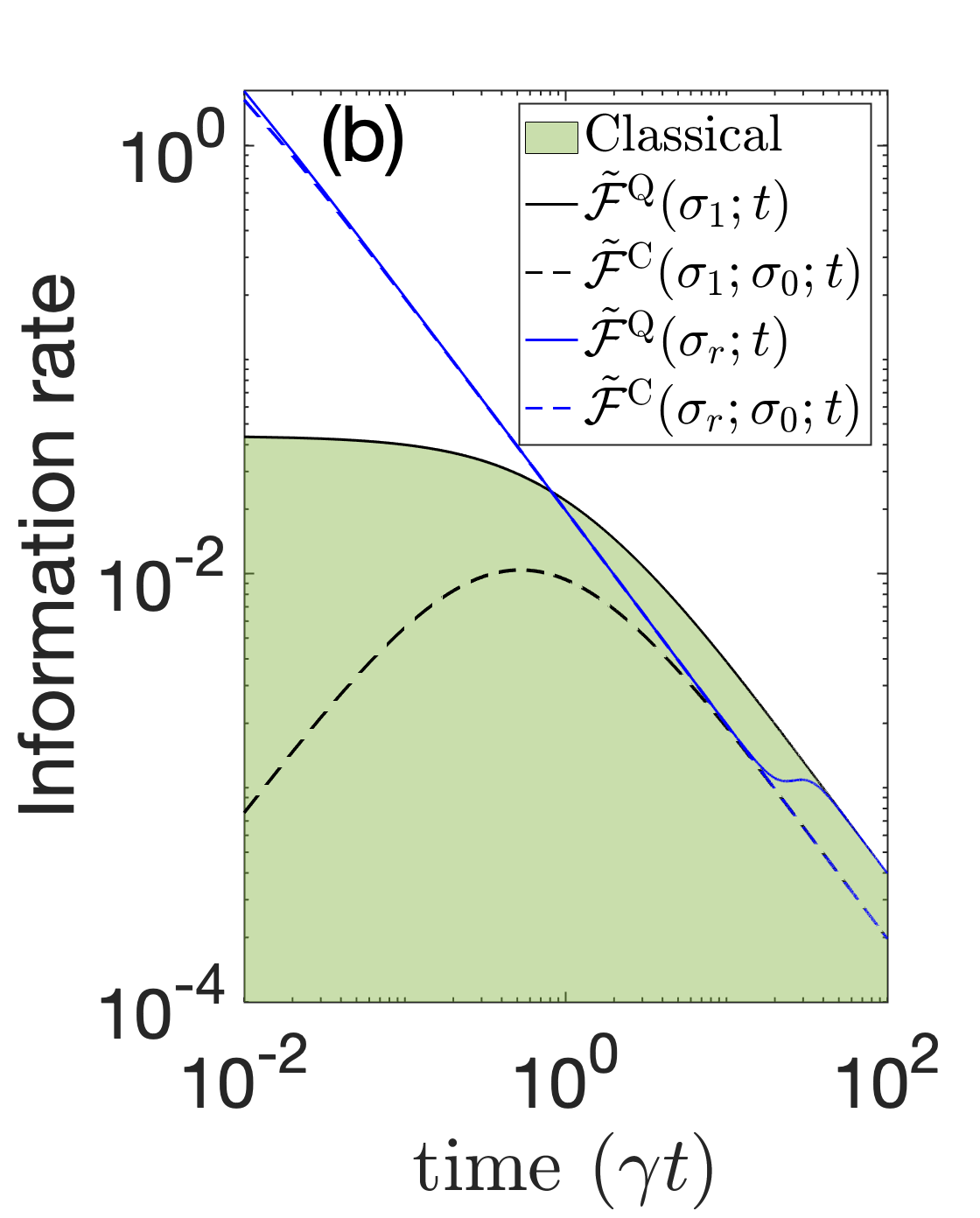}
		\caption{{The QFI rate for various temperatures, (a) $T=0.2\omega$ corresponding to the low temperature regime, and (b) $ T=5\omega$ corresponding to the high temperature regime. Similar to the main text, we observe an improvement in using squeezed probe states. Although, for smaller $T/\omega $ one observes the enhancement at smaller times. Here, we set the rest of the parameters to $\omega=1$, $\gamma =0.1$, and $r =10^{-3}$.}}
		\label{fig:app_QFI_rate_diff_T}
	\end{figure}

\begin{figure}[h!]
		\centering
		\includegraphics[width=.3\linewidth]{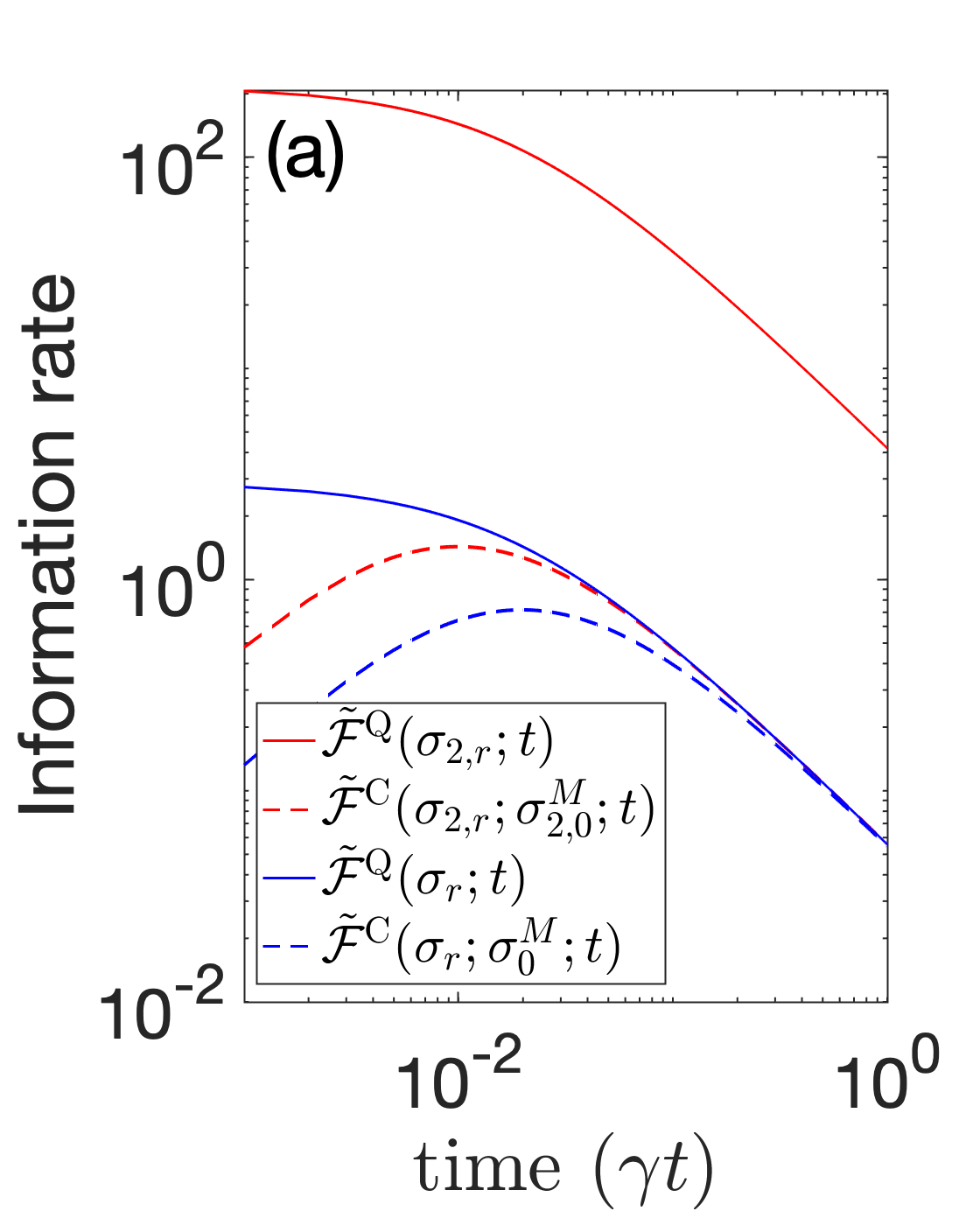}
		\includegraphics[width=.3\linewidth]{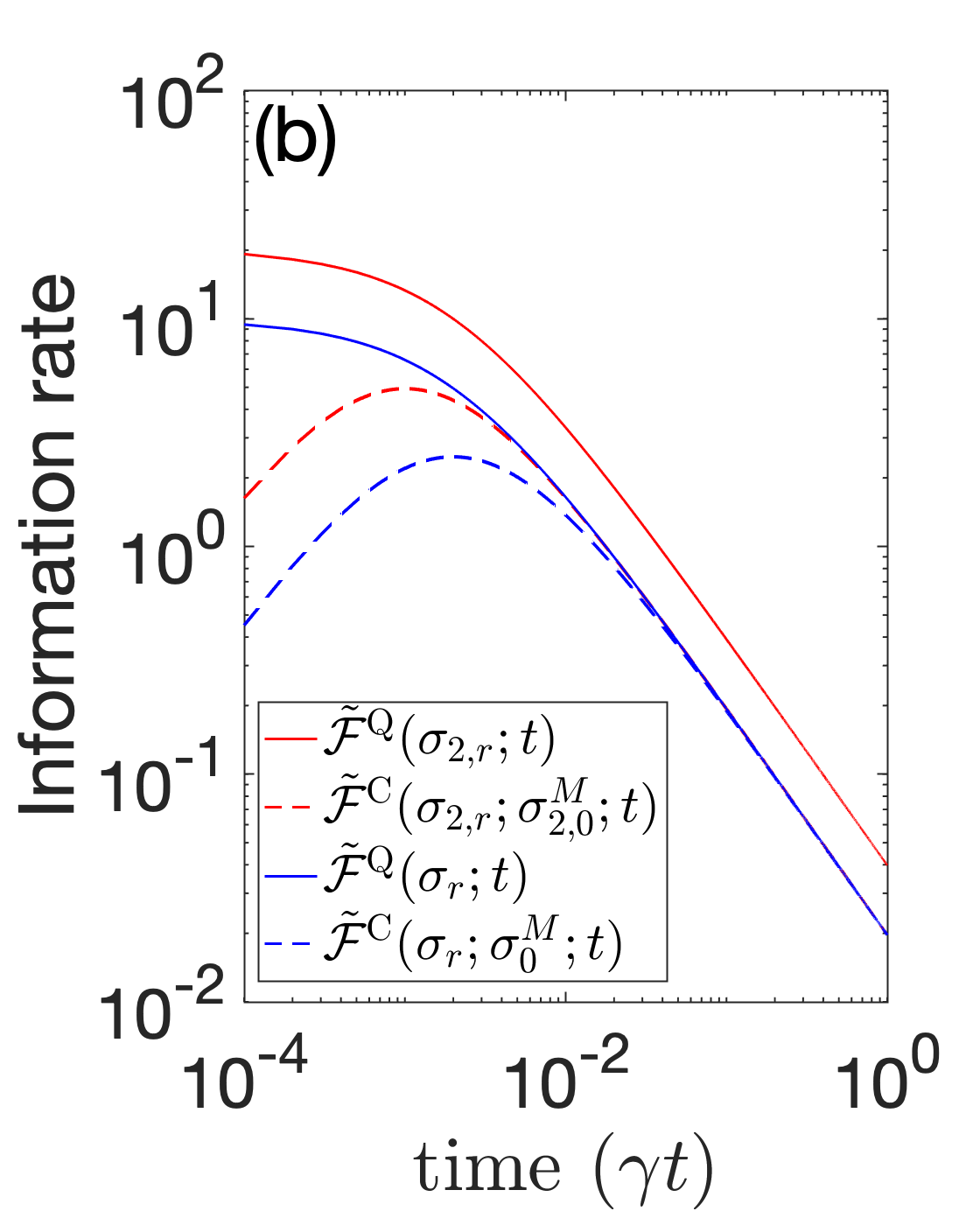}
		\caption{The QFI rate for various temperatures; comparison between two-mode and single-mode squeezed probe staets. (a) $T=0.2\omega$ corresponding to the low temperature regieme, and (b) $ T=5\omega$ corresponding to the high temperature regieme. Similar to the main text, we observe an enhancement in using entangled two-mode squeezed vacuum states. For smaller values of $T/\omega $, the enhancement can be even larger.	Here, we set the rest of the parameters to $\omega=1$, $\gamma =0.1$, and $r =10^{-3}$.}
		\label{fig:app_QFI_rate_diff_T_coupled}
	\end{figure}

    {As shown in Fig.~\eqref{fig:QFI_rates_single}, the nonclassicality of the initial probe state, in the form of squeezing, improves the QFI rate for $T=\omega$. However, this improvement is not restricted to this specific temperate. In Fig.~\ref{fig:app_QFI_rate_diff_T}, we observe a similar behavior at different values  of $T/\omega$. Moreover, as we see in Fig.~\ref{fig:app_QFI_rate_diff_T_coupled}, the two-mode squeezed vacuum states can also improve the precision at different temperatures. In particular, at lower temperatures the improvement is even more significant. Nonetheless, our ansatz  Gaussian measurement does not fully exploit this improvement.}
    

     
	\section{The optimal classical Fisher information for single-mode squeezed probe states}\label{app:opt_measure}
	
	\subsection{The covariance matrix of the optimal Gaussian measurement is always diagonal}
		As discussed in the main text, by working in the interaction picture $O_t=I_2$, the covariance matrix of the probe state, Eq.~(\ref{eq:CM-time}), becomes
	\begin{align}\label{sigmat}
		\sigma_t=e^{-\gamma t}\sigma_r+(1-e^{-\gamma t})\nu I_2.
	\end{align}
	Here, $\sigma_r={\rm diag}(r,1/r)$, as proven in the main text, is the optimal form of the initial preparation of the probe state. Thus, the covariance matrix  $\sigma_t={\rm diag}\left([\sigma_t]_{11} ,[\sigma_t]_{22}\right)$ remains diagonal. Note that, since we have symmetry in $x$- and $p$-quadrature, without loss of generality, we can assume that $0<r\leqslant1$, which implies $[\sigma_t]_{11} \leqslant [\sigma_t]_{22}$.

	To obtain the optimal single-mode Gaussian measurements, we only need to consider rank-1 measurements whose covariance matrix, in general, is described by
	\begin{align}\label{eq:singGausMeas}
		\sigma^M=\left(
		\begin{array}{cc}
			\cos\theta  & \sin\theta  \\
			-\sin\theta & \cos\theta 
		\end{array}\right)\!\left(
		\begin{array}{cc}
			L & 0 \\
			0 & L^{-1}
		\end{array}\right)\!
		\left(
		\begin{array}{cc}
			\cos\theta  & -\sin\theta  \\
			\sin\theta & \cos\theta 
		\end{array}\right)
		=\left(
		\begin{array}{cc}
			a & c \\
			c & b
		\end{array}\right)
	\end{align}
	where $0 <L \leqslant 1$ and $0\leqslant\theta<\pi$.
	Thus, two parameters $\theta$ and $r$ characterize all rank-1 Gaussian measurements. In particular, homodyne measurement on $x$ ($p$) quadrature is characterized by $L\to 0$ and $\theta=0$ ($L\to 0$ and $\theta=\pi/2$). heterodyne, the measurement in coherent state basis, is identified with $L=1$.

	The classical Fisher information (CFI) for a single-mode Gaussian probe state and Gaussian measurement, by using Eqs.~(\ref{sigmat}) and (\ref{eq:singGausMeas}), is given by
	\begin{align}
		{\cal F}^{\rm C}(\sigma_r;\sigma^M;t)&= \partial_T d_t^T \sigma_t^{-1}\partial_Td_t +
		\frac{1}{2}{\rm Tr}\! \left[\left(({\sigma}_t + {\sigma}^{M})^{-1} \partial_{T}{\sigma}_t\right)^2 \right]\\
		&=\frac{1}{2}(1-e^{-\gamma  t})^2({\partial_T \nu})^2\,{\rm Tr}\!\left[\left(
		\begin{array}{cc}
			[\sigma_t]_{11}+a & c \\
			c & [\sigma_t]_{22}+b
		\end{array}
		\right)^{\!\!-2}\right].
	\end{align}
	where we used $\partial_T d_t=0$. 
	By diagonalizing the argument,
	\begin{align}
		\left(
		\begin{array}{cc}
			[\sigma_t]_{11}+a & c \\
			c & [\sigma_t]_{22}+b
		\end{array}
		\right)^{-1}=U\left(
		\begin{array}{cc}      1/\lambda_+ & 0 \\
			0 & 1/\lambda_-
		\end{array}  
		\right)U^{\dagger},
	\end{align}
	and using the eigenvalues
	\begin{eqnarray}
		\lambda_{\pm}=\frac{1}{2}\bigg([\sigma_t]_{11}+[\sigma_t]_{22}+a+b\pm\sqrt{4c^2+\left([\sigma_t]_{11}-[\sigma_t]_{22}+a-b\right)^2}\bigg)
	\end{eqnarray}
	and Eq.~(\ref{eq:singGausMeas}), we get the general form of the CFI as follows
	\begin{align}
		{\cal F}^{\rm C}(\sigma_r;\sigma^M;t)&= \frac{1}{2}(1-e^{-\gamma  t})^2({\partial_T \nu})^2(\lambda_+^{-2}+\lambda_-^{-2})  \\
		&=2(1-e^{-\gamma  t})^2({\partial_T \nu})^2\nonumber\\
		&\times \frac{1+L\left({\rm Tr}[\sigma_t]+L(L^2+[\sigma_t]_{22}^2+[\sigma_t]_{11}^2+L{\rm Tr}[\sigma_t])\right)-L(L^2-1)\!\left([\sigma_t]_{22}-[\sigma_t]_{11}\right)\cos(2\theta)}
		{\left({\rm Tr}[\sigma_t]+L\left(2+2[\sigma_t]_{22}[\sigma_t]_{11}+L{\rm Tr}[\sigma_t]\right)+\left(L^2-1\right)\!\left([\sigma_t]_{22}-[\sigma_t]_{11}\right)\cos(2\theta)\right)^2}.\label{eq:CFIsingGausM}
	\end{align}
	There are two cases. First, for $r=1$, when the initial probe state is vacuum or coherent state, by using Eq.~(\ref{sigmat}) we have $[\sigma_t]_{22}=[\sigma_t]_{11}$. Hence, the CFI~(\ref{eq:CFIsingGausM}) is independent of $\theta$, and it can be set to zero to have a diagonal $\sigma^M$ for the Gaussian measurement; in this case, Eq.~(\ref{eq:CFIsingGausM}) becomes
	\begin{equation}
		{\cal F}^{\rm C}(\sigma_1;\sigma^M;t)=\frac{1}{2}(1-e^{-\gamma  t})^2({\partial_T \nu})^2 \bigg(\!\frac{1}{([\sigma_t]_{11}+L)^2}+\frac{1}{([\sigma_t]_{11}+L^{-1})^2}\!\bigg).
	\end{equation}
	Second, for $0<r<1$, corresponding to squeezed-vacuum probe states, we have $[\sigma_t]_{11}\leqslant[\sigma_t]_{22}$. In this case for any $0 < L\leqslant 1$, the CFI~(\ref{eq:CFIsingGausM}) is maximized when $(L^2-1)([\sigma_t]_{22}-[\sigma_t]_{11})\cos(2\theta)<0$ and $|(L^2-1)([\sigma_t]_{22}-[\sigma_t]_{11})\cos(2\theta)|$ is maximum. Hence, the optimal value is $\theta=0$, and again the covariance matrix $\sigma^M$ is diagonal; in this case, Eq.~(\ref{eq:CFIsingGausM}) becomes
	\begin{equation}\label{eq:CFI-singGaudiag}
		{\cal F}^{\rm C}(\sigma_r;\sigma^M;t)=\frac{1}{2}(1-e^{-\gamma  t})^2({\partial_T \nu})^2 \bigg(\!\frac{1}{([\sigma_t]_{11}+L)^2}+\frac{1}{([\sigma_t]_{22}+L^{-1})^2}\!\bigg).
	\end{equation}
	Therefore, in both cases, if the covariance matrix of the initial state is diagonal, the optimal rank-1 Gaussian measurement has a diagonal covariance matrix as well.

	\subsection{The optimal Gaussian measurement at short times is homodyne detection}
	
\begin{figure}
		\centering
		\includegraphics[width=.4\linewidth]{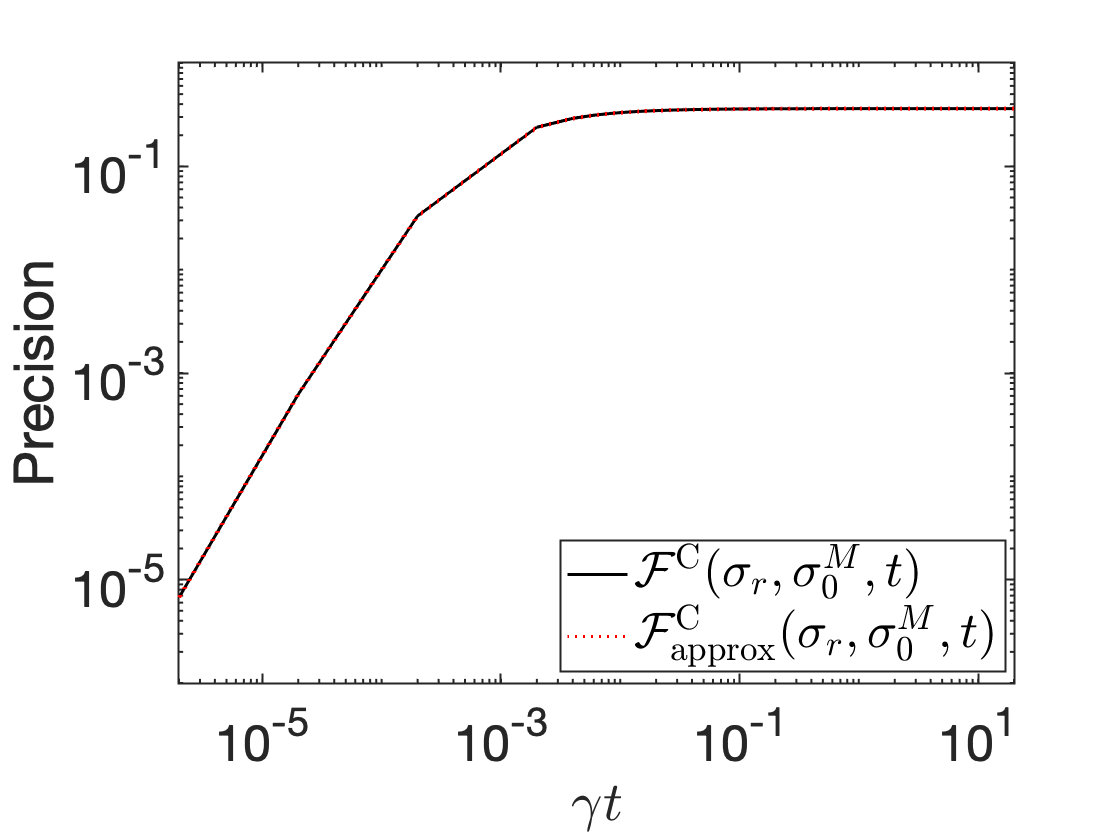}
		\includegraphics[width=.4\linewidth]{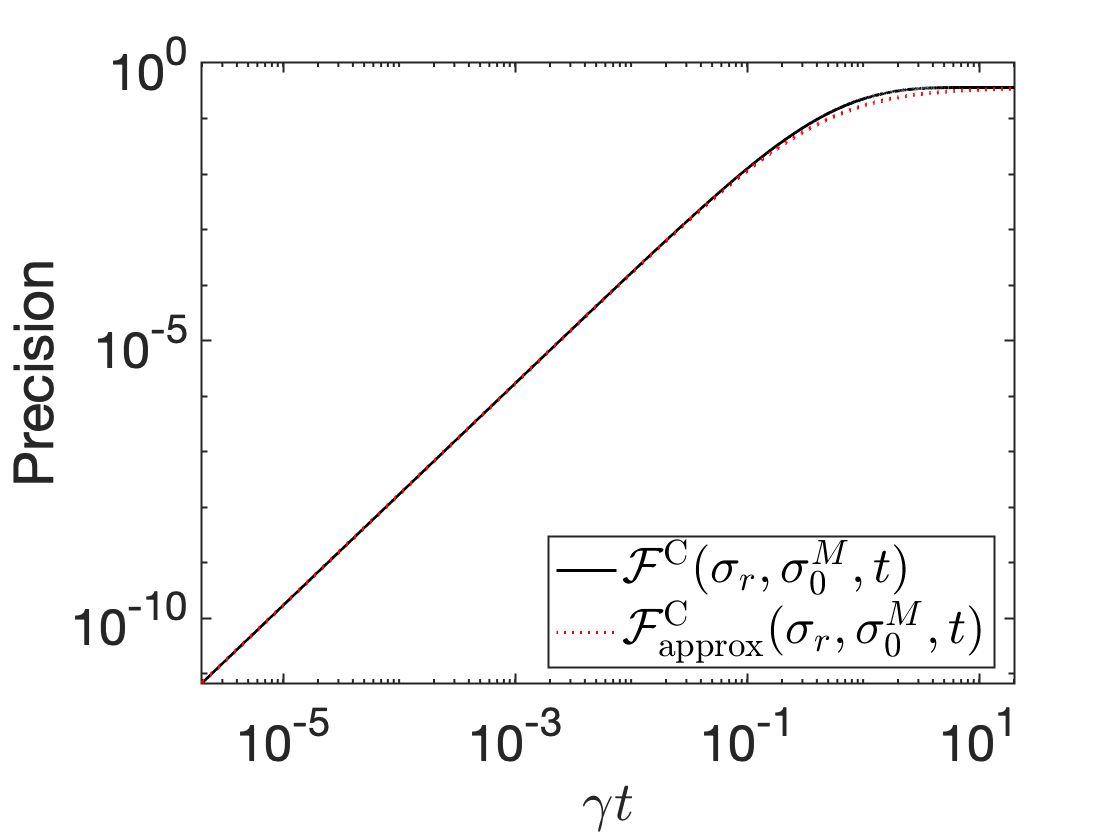}
		\caption{The comparison between the exact and the approximation of the CFI, respectively, Eqs.~\eqref{eq:CFI_Diag_exact} and \eqref{eq:CFI_Diag_approx}, shows a good agreement. In the left panel we set $r=10^{-3}$ corresponding to a highly squeezed initial preparation, while in the right panel, we set $r=1$ corresponding to vacuum. The rest of the parameters are set to $T=1$, $\gamma=0.2$, $r=10^{-3}$, and $\omega=1$. }
		\label{fig:CFI_approx}
	\end{figure}

	We start by using the CFI---Eq.~(\ref{eq:CFI-singGaudiag})---for the state given by Eq.~(\ref{sigmat}), and a Gaussian measurement with diagonal covariance matrix $\sigma^M_L={\rm diag}(L ,1/L)$. 
	The CFI reads
	\begin{align}\label{eq:CFI_Diag_exact}
		{\cal F}_{\rm exact}^{\rm C}(\sigma_r;\sigma^M_L;t )&\coloneqq {\cal F}^{\rm C}(\sigma_r;\sigma^M_L;t )=\frac{1}{2}\big(1-e^{-\gamma t}\big)^2(\partial_T \nu)^2\!\left(\!\frac{{\mathrm{e}}^{2\,\gamma \,t} }{{{\left(r-\nu +L\,{\mathrm{e}}^{\gamma \,t} +\nu \,{\mathrm{e}}^{\gamma \,t} \right)}}^2 }+\frac{L^2 \,r^2 \,{\mathrm{e}}^{2\,\gamma \,t} }{{{\left(L+r\,{\mathrm{e}}^{\gamma \,t} -L\,\nu \,r+L\,\nu \,r\,{\mathrm{e}}^{\gamma \,t} \right)}}^2 }\!\right)\nonumber\\
		& = \frac{1}{2}(1-e^{-\gamma t})^2(\partial_T \nu)^2\left(\!
		\frac{{\mathrm{e}}^{2\,\gamma \,t} \,{{\left(L+r\,{\mathrm{e}}^{\gamma \,t} -L\,\nu \,r+L\,\nu \,r\,{\mathrm{e}}^{\gamma \,t} \right)}}^2 +L^2 \,r^2 \,{\mathrm{e}}^{2\,\gamma \,t} \,{{\left(r-\nu +L\,{\mathrm{e}}^{\gamma \,t} +\nu \,{\mathrm{e}}^{\gamma \,t} \right)}}^2}{{{\left(r-\nu +L\,{\mathrm{e}}^{\gamma \,t} +\nu \,{\mathrm{e}}^{\gamma \,t} \right)}}^2 \,{{\left(L+r\,{\mathrm{e}}^{\gamma \,t} -L\,\nu \,r+L\,\nu \,r\,{\mathrm{e}}^{\gamma \,t} \right)}}^2}
		\right).
	\end{align}
	We can now expand both the numerator and the denominator for short times, and keep the leading order terms in $\gamma t$. Note that in doing so we are not allowed to ignore $\gamma t$ compared to $L$ or $r$, since they can be equally small. However, we can ignore terms like $r\gamma t$ compared to $r$, or ignore $L\gamma t$ compared to $L$. We have 
	\begin{align}
		{\cal F}^{\rm C}(\sigma_r;\sigma^M_L;t\ll 1 ) &\approx
		\frac{(\gamma t\partial_T\nu)^2}{2}\frac{(L+r+r(1+L\nu)\gamma t)^2 + L^2r^2(r+L+(L+\nu)\gamma t)^2}{(r+L+(L+\nu)\gamma t)^2(L+r+r(1+L\nu)\gamma t)^2}\nonumber\\
		& \approx \frac{(\gamma t\partial_T\nu)^2}{2}\frac{(L+r)^2+L^2r^2(r+L+\nu\gamma t)^2}{(r+L+\nu\gamma t)^2(L+r)^2}.
	\end{align}
	To proceed further, we consider two specific cases. (i) If ${\cal O}(L+r) \lesssim {\cal O} (\nu \gamma t)$ then the second term in the numerator is irrelevant compared to the first term and therefore we have 
	\begin{align}\label{eq:CFI_Diag_small_r}
		{\cal F}^{\rm C}(\sigma_r;\sigma^M_L;t\ll 1 ) &\approx \frac{(\gamma t\partial_T\nu)^2}{2(r+L+\nu\gamma t)^2},
	\end{align}
	which clearly reaches its maximum if we chose $L=0$. For this choice, the Fisher information is maximised at $t^*\propto r/(\gamma \nu)$. (ii) If ${\cal O}(L+r) > {\cal O}(\nu \gamma t)$, then we can ignore the $\nu \gamma t$ terms in the parenthesis to get
	\begin{align}\label{eq:CFI_Diag_big_r}
		{\cal F}^{\rm C}(\sigma_r;\sigma^M_L;t\ll 1 ) &\approx \frac{(\gamma t\partial_T\nu)^2(1+L^2r^2)}{2(r+L)^2}.
	\end{align}
	By taking derivative of this expression 
	with respect to $L$, we find no solutions, i.e., the function is monotonic. A simple comparison then shows that  ${\cal F}^{\rm C}(\sigma_r;\sigma^M_0;t\ll 1 ) > {\cal F}^{\rm C}(\sigma_r;\sigma^M_1;t\ll 1 )  $, that is, homodyne detection is the best Gaussian measurement in this case too. For this measurement, the Fisher information is zero initially and grows quadratically with time---until $ {\cal O}(r) \lesssim {\cal O}(\nu \gamma t)$ when we have to use the expression~\eqref{eq:CFI_Diag_small_r}. Finally, note that when $L=0$, one can use equation \eqref{eq:CFI_Diag_small_r} which covers both cases
	\begin{align}\label{eq:CFI_Diag_approx}
		{\cal F}^{\rm C}_{\rm approx}(\sigma_r;\sigma^M_0;t\ll 1 ) \coloneqq \frac{(\gamma t\partial_T\nu)^2}{2(r+\nu\gamma t)^2}.
	\end{align}
	As depicted in Fig.~\ref{fig:CFI_approx}, this approximation is in a very good agreement with the exact value of the CFI given by Eq.~\eqref{eq:CFI_Diag_exact}.
	
	
	\section{Two-mode squeezed states and a feasible joint Gaussian measurement}\label{app:feasible_meas}
	\begin{figure}[H]
		\centering
		\includegraphics[width=.7\linewidth]{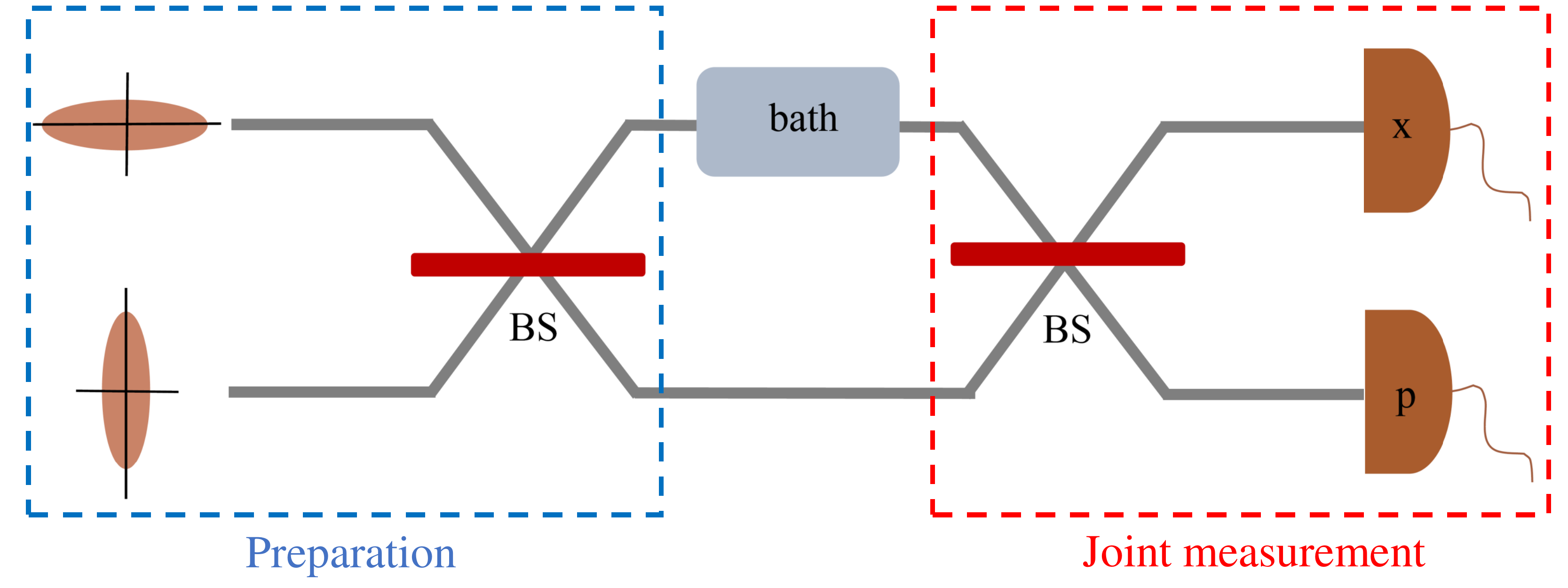}
			\caption{The schematic of a fully Gaussian setup that uses an auxiliary mode in an entangled state with probe mode and a joint measurement for further improving the thermometry precision. The dashed blue box on the left describes the preparation procedure: A two-mode squeezed vacuum state is prepared by injecting two orthogonally squeezed-vacuum states into a 50:50 beamsplitter. The probe (upper) mode interacts with the bath and then both modes are jointly measured. The joint measurement (the dashed red box) consists of a 50:50 beamsplitter and two homodyne measurements, one on the $x$-quadrature and the other on the $p$-quadrature.	\label{fig:2mode_Gaussian_meas}}
	\end{figure}	
 We describe a fully Gaussian setup that can exploit entanglement for enhanced precision in temperature estimation---see Fig.~\ref{fig:2mode_Gaussian_meas}. The joint system of probe and an auxiliary bosonic mode is prepared in a two-mode squeezed vacuum state, which is an entangled state. Such state can be generated by a spontaneous parametric down-conversion source or equivalently, as shown in Fig.~\ref{fig:2mode_Gaussian_meas}, by overlapping two single-mode squeezed vacuum states, which are orthogonally squeezed, on a 50:50 beamsplitter. By using the symplectic transformation describing the 50:50 beamsplitter which reads
	\begin{equation}
		S_{\rm BS} =\frac{1}{\sqrt2} \left(
		\begin{array}{cccc}
			1&0&1&0\\
			0&1&0&1\\
			-1&0&1&0\\
			0&-1&0&1
		\end{array}
		\right),
	\end{equation} 
	the covariance matrix of two-mode squeezed vacuum states can be obtained as
	\begin{align}
		\sigma_{2,r} =\frac{1}{2}S_{\rm BS}
		\left(
		\begin{array}{cccc}
			u&0&0&0\\
			0&u^{-1}&0&0\\
			0&0&u^{-1
			}&0\\
			0&0&0&u
		\end{array}
		\right)
		S_{\rm BS}^T =
		\left(
		\begin{array}{cc}
			r^{-1}I_2    & \sqrt{r^{-2}-1}Z\\
			\sqrt{r^{-2}-1}&
			r^{-1}I_2
		\end{array}
		\right),
	\end{align}
	where  $Z={\rm diag}(1, -1)$ and we defined $r\coloneqq 2(u+1/u)^{-1}$. After preparing the probe and the auxiliary systems in an entangled state, the probe mode interacts with the bath and, as discussed in the main text, this transforms the state into a two-mode Gaussian state that is described by the covariance matrix
	\begin{eqnarray}\label{eq:CM-2MSBath}
		\sigma_t=
		\left(
		\begin{array}{cccc}
			[(1-e^{-\gamma  t})\nu+e^{-\gamma  t}r^{-1}]I_2  & e^{-\gamma  t/2}\sqrt{r^{-2}-1}Z\\
			e^{-\gamma  t/2}\sqrt{r^{-2}-1}Z & r^{-1}I_2,
		\end{array}
		\right).
	\end{eqnarray}
	
	Now, to exploit the power of entanglement, a joint measurement is required. Our joint measurement consists of a 50:50 beamsplitter followed by two homodyne measurements, as shown in Fig.~\ref{fig:2mode_Gaussian_meas}. However, to obtain the Fisher information, we apply the beamsplitter on the state, which gives the covariance matrix $S_{\rm BS}\sigma_t S_{\rm BS} = \sigma^{\prime}_t$ with the following elements,
	\begin{align}
		[\sigma_t^{\prime}]_{11}&=[\sigma_t^{\prime}]_{44}=\frac{1-e^{-\gamma t}}{2}\nu+ \frac{1+e^{-\gamma t}}{2}r^{-1}-e^{-\gamma t/2}\sqrt{r^{-2}-1},\\ \nonumber
		[\sigma_t^{\prime}]_{13}&=[\sigma_t^{\prime}]_{24}=e^{-\gamma t/2}\sqrt{r^{-2}-1},\\ \nonumber
		[\sigma_t^{\prime}]_{22}&=[\sigma_t^{\prime}]_{33}\frac{1-e^{-\gamma t}}{2}\nu+ \frac{1+e^{-\gamma t}}{2}r^{-1}+e^{-\gamma t/2}\sqrt{r^{-2}-1},\\ \nonumber
		[\sigma_t^{\prime}]_{31}&= [\sigma_t^{\prime}]_{42}=   \frac{1-e^{-\gamma t}}{2}(\nu-r^{-1})\sqrt{r^{-2}-1},\\ 
		[\sigma_t^{\prime}]_{jk} &= 0, \text{  otherwise}.
	\end{align}
	Then, using two homodyne measurements, we measure the $x$-quadrature of the first output mode and the $p$-quadrature of the second output mode. Note that by performing this joint measurement (50:50 beamsplitter followed by two homodyne measurements), we effectively sample from a two-dimensional Gaussian probability distribution with the covariance matrix $\tilde{\sigma}={\rm diag}([\sigma_t^{\prime}]_{11},[\sigma_t^{\prime}]_{44})=[\sigma_t^{\prime}]_{11}I_2$. Therefore, the CFI is given by
	\begin{align}
		{\cal F}^{\rm C}(\sigma_{2,r};\sigma^M_{2,0};t)&=
		\frac{1}{2}{\rm Tr}\! \left[\left(\tilde{\sigma}^{-1} \partial_{T}\tilde{\sigma}\right)^2 \right]=\frac{(1-e^{-\gamma  t})^2({\partial_T \nu})^2}{4[\sigma_t^{\prime}]_{11}^2},
	\end{align}
	which is Eq.~(13) of the main text.


	\bibliography{Refs}

\end{document}